\documentclass[12pt,journal,onecolumn,draftclsnofoot]{IEEEtran}
\usepackage{graphicx,epsfig}
\usepackage{subfigure}
\usepackage{cite}
\usepackage{amsmath}
\usepackage{amssymb}
\usepackage{amsfonts,helvet}
\usepackage{footmisc}
\usepackage[alwaysadjust]{paralist}
\usepackage{algorithm}
\usepackage{algorithmic}
\usepackage{url}

\newtheorem{theorem}{Theorem}

\newtheorem{corollary}{Corollary}

\newtheorem{example}{Example}

\newtheorem{lemma}{Lemma}

\newtheorem{property}{Property}
\newtheorem{proposition}{Proposition}
\newtheorem{remark}{Remark}

\newtheorem{assumption}{\textbf{AS}}

\begin{document}
\author{\authorblockN{Vikram Chandrasekhar, Jeffrey G. Andrews, Tarik Muharemovic, Zukang Shen and Alan Gatherer} \\
\thanks{This research has been supported by Texas Instruments Inc.
V. Chandrasekhar and J. Andrews are with the Wireless Networking
and Communications Group, Dept. of Electrical and Computer
Engineering at the University of Texas at Austin, TX 78712-1157.
A. Gatherer, T. Muharemovic and Z. Shen are with Texas Instruments, Dallas, TX.
(email:cvikram@mail.utexas.edu,jandrews@ece.utexas.edu), Date: \today.}}

\title{Power Control in Two-Tier Femtocell Networks}

\maketitle

\begin{abstract}
In a two tier cellular network -- comprised of a central macrocell underlaid with shorter range femtocell hotspots -- cross-tier interference limits overall capacity with universal frequency reuse.  To quantify near-far effects with universal frequency reuse, this paper derives a fundamental relation providing the largest feasible cellular Signal-to-Interference-Plus-Noise Ratio (SINR), given any set of feasible femtocell SINRs. We provide a link budget analysis which enables simple and accurate performance insights in a two-tier network. A distributed utility-based SINR adaptation at femtocells is proposed in order to alleviate cross-tier interference at the macrocell from cochannel femtocells. The Foschini-Miljanic (FM) algorithm is a special case of the adaptation. Each femtocell maximizes their individual utility consisting of a SINR based reward less an incurred cost (interference to the macrocell).
Numerical results show greater than $30 \%$ improvement in mean femtocell SINRs relative to FM. In the event that cross-tier interference prevents a cellular user from obtaining its SINR target, an algorithm is proposed that reduces transmission powers of the strongest femtocell interferers. The algorithm ensures that a cellular user achieves its SINR target even with $100$ femtocells/cell-site, and requires a worst case SINR reduction of only $16 \%$ at femtocells. These results motivate design of power control schemes requiring minimal network overhead in two-tier networks with shared spectrum.
\end{abstract}

\section{Introduction}
Wireless operators are in the process of augmenting the macrocell network with supplemental infrastructure such as microcells, distributed antennas and relays. An alternative with lower upfront costs is to improve indoor coverage and capacity using the concept of \emph{end-consumer} installed femtocells or home base stations\cite{ChandrasekharMag2008}. A femtocell is a low power, short range ($10-50$ meters) wireless data access point (AP) that provides in-building coverage to home users and transports the user traffic over the internet-based IP backhaul such as cable modem or DSL. Femtocell users experience superior indoor reception and can lower their transmit power. Consequently, femtocells provide higher spatial reuse and cause less interference to other users.

Due to cross-tier interference in a two-tier network with shared spectrum, the target per-tier SINRs among macrocell and femtocell users are coupled. The notion of a SINR ``target" models a certain application dependent minimum
Quality of Service (QoS) requirement per user.
It is reasonable to expect that femtocell users and cellular users seek different SINRs (data rates) -- typically higher data rates using femtocells -- because home users deploy femtocells in their self interest, and because of the proximity to their BS. However, the QoS improvement arising from femtocells should come at an expense of reduced cellular coverage.

\subsection{Managing Cross-Interference in a Two-tier Network}
Contemporary wireless systems employ power control to assist users experiencing poor channels and to limit interference caused to neighboring cells. In a two-tier network however, cross-tier interference may significantly hinder the performance of conventional power control schemes. For example, signal strength based power control (channel inversion) employed by cellular users results in unacceptable deterioration of femtocell SINRs\cite{ChandrasekharCDMA2009}. The reason is because a user on its cell-edge transmits with higher power to meet its receive power target, and causes excessive cross-tier interference at nearby femtocells.

Interference management in two-tier networks faces practical challenges from the lack of coordination between the macrocell base-station (BS) and femtocell APs due to reasons of scalability, security and limited availability of backhaul bandwidth\cite{ZemlianovInfocomm2005}. From an infrastructure or spectrum availability perspective, it may be easier to operate the macrocell and femtocells in a common spectrum; at the same time, pragmatic solutions are necessary to reduce cross-tier interference. An open access (OA) scheme\cite{Claussen2007}, which performs radio management by vertical handoffs -- forcing cellular users to communicate with nearby femtocells to load balance traffic in each tier -- is one such solution. A drawback of OA is the network overhead \cite{ChandrasekharMag2008,Niyato2005} and the need for sufficient backhaul capacity to avoid starving the paying home user. Additionally, OA potentially compromises security and QoS for home users.

This work assumes \emph{Closed Access} (CA), which means only licensed home users within radio range can communicate with their own femtocell. With CA, cross-tier interference from interior femtocells may significantly deteriorate the SINR at the macrocell BS. The motivation behind this paper is ensuring that the service (data rates) provided to cellular users remain unaffected by a femtocell underlay which operates in the same spectrum. Three main reasons are \begin{inparaenum} \item the macrocell's primary role of an anytime anywhere infrastructure, especially for mobile and ``isolated" users without hotspot access, \item the greater number of users served by each macrocell BS, and \item the end user deployment of femtocells in their self-interest\end{inparaenum}.  The macrocell is consequently modeled as primary infrastructure, meaning that the operator's foremost obligation is to ensure that an outdoor cellular user achieves its minimum SINR target at its BS, despite cross-tier femtocell interference. Indoor users act in their self interest to maximize their SINRs, but incur a SINR penalty because they cause cross-tier interference.

Considering a macrocell BS with $N$ cochannel femtocells and one transmitting user per slot per cell over the uplink, the following questions are addressed in this paper:
\begin{itemize}
\item Given a set of feasible target SINRs inside femtocell hotspots, what is the largest cellular SINR target for which a non-negative power allocation exists for all users in the system?
\item How does the cellular SINR depend on the locations of macrocell and femtocell users and cellular parameters such as the channel gains between cellular users and femtocells?
\item Given an utility-based femtocell SINR adaptation with a certain minimum QoS requirement at each femtocell, what are the ensuing SINR equilibria and can they be achieved in a distributed fashion?
\item When a cellular user cannot satisfy its SINR target due to cross-tier interference, by how much should femtocells reduce their SINR target to ensure that the cellular user's SINR requirement is met?
\end{itemize}
Although this work exclusively focuses on the uplink in a tiered cellular system, we would like to clarify that portions of our analysis (Section III) are also applicable in the downlink with potentially different conclusions. Due to space limitations, the downlink extension is omitted for future work.


\subsection{Prior Work}
Prior research in cellular power control and rate assignments in tiered networks mainly considered an operator planned underlay of a macrocell with single/multiple microcells\cite{Ganz1997,Kishore2005}. In the context of this paper, a microcell has a much larger radio range (100-500 m) than a femtocell, and generally implies centralized deployment, i.e. by the service-provider. A microcell underlay allows the operator to handoff and load balance users between each tier\cite{ChandrasekharMag2008}. For example, the operator can preferentially assign high data rate users to a microcell \cite{Klein2004,Kishore2005,Shen2004} because of its inherently larger capacity. In contrast, femtocells are consumer installed and the traffic requirements at femtocells are user determined without any operator influence. Consequently, distributed interference management strategies may be preferred.

Our work ties in with well known power control schemes in conventional cellular networks and prior work on utility optimization based on game theory. Results in Foschini \emph{et al.}\cite{Foschini1993}, Zander\cite{Zander1992a}, Grandhi \emph{et al.}\cite{Grandhi1994} and Bambos \emph{et al.}\cite{Bambos2000} provide conditions for SINR feasibility and/or SIR balancing in cellular systems. Specifically, in a network with $N$ users with target SINRs $\Gamma_i, 1 \leq i \leq N$, a feasible power allocation for all users exists iff the spectral radius of the normalized channel gain matrix is less than unity. Associated results on centralized/distributed/constrained power control, link admission control and user-BS assignment are presented in \cite{Zander1992b,Yates1995a,Yates1995b,Hanly1995,Ulukus1998,Grandhi1994,Sung2005} and numerous other works.

The utility-based non-cooperative femtocell SINR adaptation presented here is related to existing game theory literature on non-cooperative cellular power control\cite{JiHuang1998,Goodman2000,Saraydar2002,Koskie2005,Xiao2001,Alpcan2001} (see \cite{Altman2006} for a survey). The adaptation forces stronger femtocell interferers to obtain their SINR equilibria closer to their minimum SINR targets, while femtocells causing smaller cross-tier interference obtain higher SINR margins. This is similar to Xiao and Shroff\cite{Xiao2001}'s utility-based power control (UBPC) scheme, wherein users vary their target SIRs based on the prevailing traffic conditions. Unlike the sigmoidal utility in \cite{Xiao2001}, our utility function has a more meaningful interpretation because it models \begin{inparaenum} \item the femtocell user's inclination to seek higher data-rates and \item the primary role of the macrocell while penalizing the femtocell user for causing cross-tier interference. \end{inparaenum} Our SINR equilibria is simple to characterize unlike the feasibility conditions presented in prior works e.g \cite{Alpcan2001}.

To minimize cross-tier interference, prior femtocell research has proposed open access \cite{Claussen2007}, varying femtocell coverage area \cite{Ho2007}, hybrid frequency assignments\cite{Guvenc2008}, adjusting the maximum transmit power of femtocell users \cite{Jo2008} and adaptive access operation of femtocells \cite{Choi2008}. In contrast, this paper addresses SINR adaptation and ensuring acceptable cellular performance in closed access femtocells. Related works in cognitive radio (CR) literature such as \cite{Qian2007,Hoven2005} propose that secondary users limit their transmission powers for reducing interference to primary users (PUs).
In \cite{Hoven2005}, CR users regulate their transmit powers to limit PU interference, but their work does not address individual rate requirements at each CR. Qian \emph{et al.}\cite{Qian2007} propose a joint power and admission control scheme, but provide little insight on how a CR user's data-rate is influenced by a PU's rate. In contrast, our results are applicable in CR networks for determining the \emph{exact relationship} between the feasible SINRs of primary and CR users; further our SINR adaptation can enable CR users to vary their data-rates in a decentralized manner based on instantaneous interference at PU receivers.

\subsection{Contributions}
\begin{asparadesc}
\item[Pareto SINR Contours.] Near-far effects in a cochannel two-tier network are captured through a theoretical analysis providing the highest cellular SINR target--for which a non-negative power allocation exists between all transmit-receive pairs--given any set of femtocell SINRs and vice versa. With a common SINR target at femtocells and neglecting interference between femtocells, the per-tier Pareto SINR pairs have an intuitive interpretation: the sum of the decibel (dB) cellular SINR and the dB femtocell SINR equals a constant. Design interpretations are provided for different path loss exponents, different numbers of femtocells and varying locations of the cellular user and hotspots.
\item[Utility-based Femtocell SINR Adaptation.] Femtocells individually maximize an objective function consisting of a SINR dependent reward, and a penalty proportional to the interference at the macrocell. We obtain a \emph{channel-dependant SINR equilibrium} at each femtocell. The equilibrium discourages strongly interfering femtocells to use large transmit powers. This SINR equilibrium is attained using distributed power updates\cite{Yates1995b}. For femtocell users whose objective is to simply equal their minimum SINR targets, our adaptation simplifies to the Foschini-Miljanic (FM) update. Numerical results show that the utility adaptation provides up to $30 \%$ higher femtocell SINRs relative to FM.
\item [Cellular Link Quality Protection.] To alleviate cross-tier interference when the cellular user does not achieve its SINR target, we propose a distributed algorithm to progressively reduce SINR targets of strongest femtocell interferers until the cellular SINR target is met. Numerical simulations with $100$ femtocells/cell-site show acceptable cellular coverage with a worst-case femtocell SINR reduction of only $16 \%$ (with typical cellular parameters).

\end{asparadesc}

\section{System Model}
The system consists of a single central macrocell $B_0$ serving a region $\mathcal{C}$, providing a cellular coverage radius $R_c$. The macrocell is underlaid with $N$ cochannel femtocells APs $B_i,i \geq 1$. Femtocell users are located on the circumference of a disc of radius $R_f$ centered at their femtocell AP. Orthogonal uplink signaling is assumed in each slot ($1$ scheduled active user per cell during each signaling slot), where a slot may refer to a time or frequency resource (the ensuing analysis leading up to Theorem \ref{th:ParetoSINRContour} apply equally well over the downlink).

\begin{assumption}
\label{AS:CCI}
For analytical tractability, cochannel interference from neighboring cellular transmissions is ignored.
\end{assumption}

During a given slot, let $i \in \{0,1,\cdots, N\}$ denote the scheduled user connected to its BS $B_i$. Designate user $i$'s transmit power to be $p_i$ Watts. Let $\sigma^2$ be the variance of Additive White Gaussian Noise (AWGN) at $B_i$. The received SINR $\gamma_i$ of user $i$ at $B_i$ is given as
\begin{align}
\label{eq:Twotier_SINR}
\Gamma_i \leq \gamma_i=\frac{p_i g_{i,i}}{\sum_{j \neq i} p_j g_{i,j}+\sigma^2}.
\end{align}
Here $\Gamma_i$ represents the minimum target SINR for user $i$ at $B_i$. The term $g_{i,j}$ denotes the channel gain between user $j$ and BS $B_i$. Note that $g_{i,i}$ can also account for post-processing SINR gains arising from, but not restricted to, diversity reception or interference suppression (e.g. CDMA). In matrix-vector notation, \eqref{eq:Twotier_SINR} can be written as
\begin{align}
\label{eq:Twotier_SINRVector}
\mathbf{p} \geq \mathbf{\Gamma}\mathbf{G}\mathbf{p}+\boldsymbol{\eta} \textrm{ and } \mathbf{p} \geq 0.
\end{align}
Here $\mathbf{\Gamma} \triangleq \mathrm{diag}(\Gamma_0, \dots \Gamma_N)$ while the vector $\mathbf{p}=(p_0,p_1,\cdots p_{N})$ denotes the transmission powers of individual users, and the normalized noise vector equals $\boldsymbol{\eta}=(\eta_0, \dots \eta_{N}),\eta_i=\sigma^2 \Gamma_i/g_{i,i}$. The $(N+1) \times (N+1)$ matrix $\mathbf{G}\geq 0$ is assumed to be irreducible -- meaning its directed graph is strongly connected \cite[Page 362]{Horn1985} -- with elements given as
\begin{align}
G_{ij}=\frac{g_{i,j}}{g_{i,i}}, i \neq j \textrm{ and} \ 0  \textrm{ else}.
\end{align}
Since $\mathbf{\Gamma G}$ is nonnegative, the spectral radius $\rho(\mathbf{\Gamma G})$ (defined as the maximum modulus eigenvalue $\max \lbrace |\lambda|:\mathbf{\Gamma G}-\lambda \mathbf{I}_{N+1} \textrm{ is singular}\rbrace$) is an eigenvalue of $\mathbf{\Gamma G}$ \cite[Theorem 8.3.1]{Horn1985}. Applying Perron-Frobenius theory \cite{Horn1985} to $\mathbf{\Gamma}\mathbf{G}$, \eqref{eq:Twotier_SINRVector} has a nonnegative solution $\mathbf{p}^{\ast}$ (or $\mathbf{\Gamma}$ constitutes a \emph{feasible} set of target SINR assignments) \emph{iff} the spectral radius $\rho(\mathbf{\Gamma}\mathbf{G})$ is less than unity\cite{Grandhi1994,Bambos2000}. Consequently,
\begin{align}
\label{eq:Twotier_PFcondition}
\forall \boldsymbol{\eta} \geq \mathbf{0}, (\mathbf{I}-\mathbf{\Gamma}\mathbf{G})^{-1}>0 \Leftrightarrow (\mathbf{I}-\mathbf{\Gamma}\mathbf{G})^{-1}\boldsymbol{\eta} \geq \mathbf{0} \Leftrightarrow \rho(\mathbf{\Gamma}\mathbf{G})<1.
\end{align}
The solution $\mathbf{p}^{\ast}=(\mathbf{I}-\mathbf{\Gamma}\mathbf{G})^{-1}\boldsymbol{\eta}$ guarantees that the target SINR requirements are satisfied at all BSs. Further, $\mathbf{p}^{\ast}$ is Pareto efficient in the sense that any other solution $\mathbf{p}$ satisfying \eqref{eq:Twotier_SINRVector} needs at least as much power componentwise\cite{Bambos2000}.  When $\mathbf{\Gamma}=\gamma \mathbf{I}_{N+1}$, then the max-min SIR solution $\gamma^{\ast}$ to \eqref{eq:Twotier_PFcondition} is given as
\begin{align}
\label{eq:Zander}
\mathbf{\Gamma}=\Gamma \mathbf{I}_{N+1} \Rightarrow \Gamma^{\ast}=\frac{1}{\rho(\mathbf{G})}.
\end{align}
In an interference-limited system (neglecting $\boldsymbol{\eta}$), the optimizing vector $\mathbf{p}^{\ast}$ equals the Perron-Frobenius eigenvector of $\mathbf{\Gamma}\mathbf{G}$ \cite{Zander1992a}.

\section{Per-Tier SINR Contours In a Femtocell-Underlaid Macrocell}
In a two-tier network, let $\Gamma_c = \Gamma_0$ and $\Gamma_i \  (i \geq 1)$ denote the per-tier SINR targets at the macrocell and femtocell BSs respectively. Define $\mathbf{\Gamma}_f \triangleq \mathrm{diag}(\Gamma_1, \Gamma_2, \dots,\Gamma_N)$ and $\mathbf{\Gamma} = \mathrm{diag}(\Gamma_c,\mathbf{\Gamma}_f)$. Any feasible SINR tuple ensures that the spectral radius $\rho(\mathbf{\Gamma G}) <1$ with a feasible power assignment given by \eqref{eq:Twotier_PFcondition}. This section derives the relationship between $\Gamma_c$ and $\Gamma_i$ as a function of $\kappa$ and entries of the $\mathbf{G}$ matrix.

Using the above notation, $\mathbf{\Gamma G}$ simplifies as
\begin{align}
\label{eq:GMtxDecomp}
\mathbf{\Gamma G}= \begin{pmatrix}
                                0 & \Gamma_c \mathbf{q}_c^{T} \\
                                \mathbf{\Gamma}_f \mathbf{q}_f & \mathbf{\Gamma}_f \mathbf{F}
                        \end{pmatrix}.
\end{align}
Here the principal submatrix $\mathbf{F}$ consists of the normalized channel gains between each femtocell and its surrounding $N-1$ cochannel femtocells. The vector $\mathbf{q}_C^{T} = \lbrack G_{01},G_{02}, \dots, G_{0N}\rbrack$ consists of the normalized cross-tier channel gains between the transmitting femtocell users to the macrocell BS. Similarly, $\mathbf{q}_F = \lbrack G_{10}, G_{20}, \dots G_{N0} \rbrack^{T}$ consists of the normalized cross-tier channel gains between the cellular user to surrounding femtocell BSs.

Below, we list two simple but useful properties of $\mathbf{\Gamma G}$:
\begin{property}
\label{prop:P1}
\emph{
$\rho(\mathbf{\Gamma G})$ is a non-decreasing function of $\mathbf{\Gamma}$. That is, $\mathbf{\Gamma}' \geq \mathbf{\Gamma} \Rightarrow \rho(\mathbf{\Gamma}' \mathbf{G}) \geq \rho(\mathbf{\Gamma} \mathbf{G})$}.
\end{property}
\begin{property}
\label{prop:P2}
\emph{
$\rho(\mathbf{\Gamma G}) \geq \rho(\mathbf{\Gamma}_f \mathbf{F})$}.
\end{property}
Property \ref{prop:P1} is a consequence of \cite[Corollary 8.1.19]{Horn1985} and implies that increasing the per-tier SINRs in $\mathbf{\Gamma}$ drives $\rho(\mathbf{\Gamma G})$ closer to unity. This decreases the margin for existence of a nonnegative inverse of $\mathbf{I}-\mathbf{\Gamma G}$ in \eqref{eq:Twotier_PFcondition}. Therefore, assuming a fixed set of femtocell SINRs given by $\mathbf{\Gamma}_f$, the maximum cellular SINR target $\Gamma_0$ monotonically increases with $\rho(\mathbf{\Gamma G})$. Property \ref{prop:P2} arises as a consequence of $\mathbf{\Gamma}_f \mathbf{F}$ being a principal submatrix of $\mathbf{G}$, and applying \cite[Corollary 8.1.20]{Horn1985}. Intuitively, any feasible femtocell SINR in a tiered network is also feasible when the network comprises only femtocells since $\rho(\mathbf{\Gamma G}) < 1 \Rightarrow \rho(\mathbf{\Gamma}_f \mathbf{F}) <1$. From \eqref{eq:Twotier_PFcondition}, the condition $ \rho(\mathbf{\Gamma}_f \mathbf{F}) <1 \Leftrightarrow (\mathbf{I}-\mathbf{\Gamma}_f \mathbf{F})^{-1}$ is nonnegative with expansion given as $\sum_{k=0}^{\infty}(\mathbf{\Gamma}_f \mathbf{F})^k$.

We restate a useful lemma by Meyer\cite{Meyer1989} for obtaining $\rho(\mathbf{\Gamma G})$ in terms of $\mathbf{F},\mathbf{q}_f,\mathbf{q}_c,\Gamma_c$ and $\mathbf{\Gamma}_f$.
\begin{lemma}\label{le:Meyer}
\cite[Meyer]{Meyer1989}
\emph{Let $\mathbf{A}$ be a $m \times n$ nonnegative irreducible matrix with spectral radius $\rho$ and let $\mathbf{A}$ have a k-level partition
\begin{align}
\mathbf{A} = \begin{pmatrix} \mathbf{A}_{11} & \mathbf{A}_{12} & \dots & \mathbf{A}_{1k} \\
                    \mathbf{A}_{21} & \mathbf{A}_{22} & \dots & \mathbf{A}_{2k} \\
                    \vdots        & \vdots        & \ddots & \vdots \\
                    \mathbf{A}_{k1} & \mathbf{A}_{k2} & \dots & \mathbf{A}_{kk}
    \end{pmatrix}
\end{align}
 in which all diagonal blocks are square. For a given index $i$, let $\mathbf{A}_i$ represent the principal block submatrix of $\mathbf{A}$ by deleting the $i$th row and $i$th column of blocks from $\mathbf{A}$. Let $\mathbf{A}_{i \ast}$ designate the $i$th row of blocks with $\mathbf{A}_{ii}$ removed. Similarly, let $\mathbf{A}_{\ast i}$ designate the $i$th column of blocks with $\mathbf{A}_{ii}$ removed.
 Then each Perron complement
$\mathbf{P}_{ii} = \mathbf{A}_{ii}+\mathbf{A}_{i \ast}(\rho \mathbf{I}-\mathbf{A}_i)^{-1}\mathbf{A}_{\ast i}$
is also a nonnegative matrix whose spectral radius is again given by $\rho$.
}
\end{lemma}

Using Lemma \ref{le:Meyer}, we state the first result in this paper.
\begin{theorem}
\label{th:ParetoSINRContour}\emph{
 Assume a set of feasible femtocell SINRs targets $\Gamma_i (i \geq 1)$ such that $\rho(\mathbf{\Gamma}_f \mathbf{F}) <1$, and a target spectral radius $\rho(\mathbf{\Gamma G}) = \kappa, \rho(\mathbf{\Gamma}_f \mathbf{F}) < \kappa < 1$. The highest cellular SINR target maintaining a spectral radius of $\kappa$ is then given as
\begin{align}
\label{eq:MacroSINR}
\Gamma_c = \frac{\kappa^2}{\mathbf{q}_c^{T}[\mathbf{I}-(\mathbf{\Gamma}_f/ \kappa)\mathbf{F}]^{-1}\mathbf{\Gamma}_f \mathbf{q}_f}.
\end{align}}
\end{theorem}
\vspace{2mm}
\begin{proof}
From Lemma \ref{le:Meyer}, the Perron complement of the entry ``$0$'' of $\mathbf{\Gamma G}$ in \eqref{eq:GMtxDecomp} is a nonnegative scalar equaling $\kappa$. This implies,
\begin{align}
\kappa = 0 + \Gamma_c \mathbf{q}_c^{T} [\kappa \mathbf{I}-\mathbf{\Gamma}_f \mathbf{F}]^{-1}\mathbf{\Gamma}_f \mathbf{q}_f.
\end{align}
Rearranging terms, we obtain \eqref{eq:MacroSINR}. Note that since $\kappa > \rho(\mathbf{\Gamma}_f\mathbf{F})$, the inverse $[\mathbf{I}-(\mathbf{\Gamma}_f /\kappa)\mathbf{F}]^{-1}=\sum_{k=0}^{\infty}(\Gamma_f/\kappa)^{k}\mathbf{F}^k$ exists and is nonnegative.
\end{proof}

Given a set of $N$ feasible femtocell SINR targets, Theorem \ref{th:ParetoSINRContour} provides a fundamental relationship describing the maximum SINR target at the macrocell over all power control strategies. Given a $\kappa$ (e.g. $\kappa = 1-\epsilon$, where $ 0 < \epsilon < 1-\rho(\mathbf{\Gamma}_f\mathbf{F})$), one obtains the highest $\Gamma_c$ for a given $\mathbf{\Gamma}_f$.
\begin{example}[\textbf{One Femtocell}]
 Consider a two-tier network consisting of the central macrocell $B_0$ and a single femtocell BS $B_1$. The matrix $\mathbf{\Gamma G}$ is given as
\begin{align}
\mathbf{\Gamma}\mathbf{G}=\begin{pmatrix} 0 & \Gamma_c G_{01} \\
                    \Gamma_f G_{10} & 0
        \end{pmatrix}.
\end{align}
Setting $\mathbf{F}=0, \mathbf{q}_c = G_{01}, \mathbf{q}_f = G_{10}$ in \eqref{eq:MacroSINR}, one obtains
\begin{align}
\rho(\mathbf{\Gamma G}) = \sqrt{\Gamma_cG_{01}\Gamma_fG_{10}}
\Rightarrow (\Gamma_c, \Gamma_f) \in \left\lbrace (x,y)\in \mathbb{R}_{+}^2: xy < \frac{1
}{G_{01}G_{10}} \right \rbrace.
\end{align}
Intuitively, the product of the per-tier SINR targets is limited by the inverse product of the cross-tier gains between the cellular user to the femtocell AP and vice versa.
\end{example}
\begin{remark}
Equation \eqref{eq:MacroSINR} generically applies in a wireless network with $N+1$ users for finding the best SINR target for a particular user -- by appropriately adjusting the entries in $\mathbf{q}_c$, $\mathbf{q}_f$ and $\mathbf{F}$ -- for a given set of $N$ SINR targets. However, the subsequent analysis (Lemma \ref{le:ParetoSINRContourUB}) specializes \eqref{eq:MacroSINR} to a two-tier cellular system and works only when the cellular user is isolated.
\end{remark}
With $\Gamma_c$ obtained from \eqref{eq:MacroSINR} and SINR targets $\mathbf{\Gamma}^{\ast} = \lbrack \Gamma_c, \Gamma_1, \Gamma_2 ,\dots, \Gamma_N \rbrack^{T}$, a centralized power allocation is given as
\begin{align}
\label{eq:TwoTierCentralizedPC}
 \mathbf{p}^{\ast}=(\mathbf{I}-\mathbf{\Gamma}^{\ast}\mathbf{G})^{-1}\boldsymbol{\eta}^{\ast} \textrm{, where } \boldsymbol{\eta}^{\ast}\triangleq \mathrm{diag}\left(\frac{\sigma^2}{g_{1,1}},\frac{\sigma^2}{g_{2,2}},\dots,\frac{\sigma^2}{g_{N+1,N+1}}\right)\mathbf{\Gamma}^{\ast}.
\end{align}
Next, assume that the $N$ femtocells $B_1 \dots B_N$ choose a \emph{common} SINR target $\Gamma_i = \Gamma_f (i\geq 1)$. Although the assumption of a common SINR target at all femtocells seems rather restrictive at first glance, it provides intuition on near-far effects in a two-tier network which will be discussed in the next section. The following corollary derives the Pareto contours between the best SINR targets for macrocell and femtocell users respectively.
\begin{corollary}
\label{co:ParetoSINRContour}\emph{
 Assume a common positive target femtocell SINR target $\Gamma_f<1/\rho(\mathbf{F})$, and a target spectral radius $\rho(\mathbf{\Gamma G}) = \kappa$, where $\Gamma_f \rho(\mathbf{F}) < \kappa < 1$. The Pareto contours maintaining a spectral radius of $\kappa$ are given as
\begin{align}
\label{eq:ParetoSINRContour}
\left \lbrace (\Gamma_c,\Gamma_f):0 \leq \Gamma_f < \frac{1}{\rho(\mathbf{F})},
\Gamma_c = \frac{\kappa^2}{\Gamma_f \mathbf{q}_c^{T}[\mathbf{I}-(\Gamma_f/ \kappa)\mathbf{F}]^{-1}\mathbf{q}_f} \right \rbrace.
\end{align}}
\end{corollary}
\vspace{2mm}
\begin{remark}[\textbf{Pareto optimality}]
Given a target spectral radius $\kappa$, the $(\Gamma_c, \mathbf{\Gamma}_f)$ tuples derived in \eqref{eq:MacroSINR} (and hence \eqref{eq:ParetoSINRContour}) are Pareto optimal. From Property \ref{prop:P1}, a ``better pair" $\mathbf{\Gamma}_f' \geq \mathbf{\Gamma}_f$ (component-wise) and $\Gamma_c' > \Gamma_c$ cannot be obtained without $\rho(\mathbf{\Gamma G})$ exceeding $\kappa$.
\end{remark}
\vspace{2mm}
\begin{lemma}
\label{le:ParetoSINRContourUB}
\emph{
With a set of feasible femtocell SINRs thresholds $\Gamma_i (i \geq 1)$ and $\rho(\mathbf{\Gamma}_f \mathbf{F}) <1$, a necessary condition for any cellular SINR target $\Gamma_c$ to be feasible is given as
\begin{align}
\label{eq:ParetoSINRContourUB1}
\Gamma_c  \leq \frac{1}{\mathbf{q}_c^{T} \mathbf{\Gamma}_f \mathbf{q}_f}.
\end{align}
Consequently, assuming a common positive SINR target $\Gamma_f<1/\rho(\mathbf{F})$ at femtocells ($1/\rho(\mathbf{F})$ being the max-min target), any feasible SINR pair $(\Gamma_c,\Gamma_f)$ satisfies the following inequality
\begin{align}
\label{eq:ParetoSINRContourUB2}
\Gamma_c \Gamma_f  < \frac{1}{\mathbf{q}_c^{T} \mathbf{q}_f}.
\end{align}}
\end{lemma}
\vspace{2mm}
\begin{proof}
Computing the Perron complement of $\mathbf{\Gamma}_f\mathbf{F}$ in \eqref{eq:GMtxDecomp} and applying Lemma \ref{le:Meyer}:
\begin{align}
\label{eq:ParetoSINRContourUBPf}
\kappa = \rho(\Gamma_f\mathbf{F}+\Gamma_f \mathbf{q}_f \Gamma_c \mathbf{q}_c^{T}/\kappa)
                        \overset{(b)}\geq \rho(\Gamma_f \mathbf{q}_f \Gamma_c \mathbf{q}_c^{T}/\kappa)
\end{align}
where step (b) in \eqref{eq:ParetoSINRContourUBPf} follows by applying \cite[Corollary 8.1.19]{Horn1985}. Upper bounding $\kappa^2$ by unity and applying $\rho(\mathbf{q}_f \mathbf{q}_c^{T})=\mathbf{q}_c^{T}\mathbf{q}_f$ to \eqref{eq:ParetoSINRContourUBPf} yields \eqref{eq:ParetoSINRContourUB1}. Alternatively, one can expand $\mathbf{I}-(\Gamma_f/ \kappa)\mathbf{F}$ and replace $\mathbf{q}_c^{T}[\mathbf{I}-(\Gamma_f/ \kappa)\mathbf{F}]^{-1}\mathbf{q}_f$ by the lower bound $\mathbf{q}_c^{T}\mathbf{q}_f$.
\end{proof}

Intuitively, \eqref{eq:ParetoSINRContourUB2} restates that $1/\mathbf{q}_c^{T}\mathbf{q}_f$ is an upper bound on the product of the per-tier SINRs, achieved when $\mathbf{F}=\mathbf{0}$ in \eqref{eq:MacroSINR}, i.e. the interference between neighboring femtocells is vanishingly small. Ignoring $\mathbf{F}$ is justifiable because \begin{inparaenum} \item the propagation between femtocells suffers at least a double wall partition losses (from inside a femtocell to outdoor and from outdoor onto the neighboring femtocell), and \item there is only one partition loss term while considering the propagation loss between a cellular user to femtocells. \end{inparaenum}

Thus, a simple relationship between the highest per-tier SINRs is expressed as:

\emph{For small $\mathbf{F}$, the sum of the per-tier decibel SINRs equals a channel dependant constant $L_{\textrm{dB}} = -10\log_{10}(\mathbf{q}_c^T\mathbf{q}_f)$.} We denote this constant $L = \frac{1}{\mathbf{q}_c^{T}\mathbf{q}_f}$ as the \emph{Link Budget}. Choosing a cellular SINR target of $x$ dB necessitates any feasible femtocell SINR target to be no more than $L_{\textrm{dB}}-x$ dB. To keep $L$ large, it is desirable that the normalized interference powers are decorrelated (or $\mathbf{q}_c$ and $\mathbf{q}_f$ do not peak simultaneously). In a certain sense, the link budget provides an ``efficiency index" of closed access femtocell operation, since open (or public) femtocell access potentially allows users to minimize their interference by handoffs.

\begin{figure} [htp]
\begin{center}
   \includegraphics[width=4.0in]{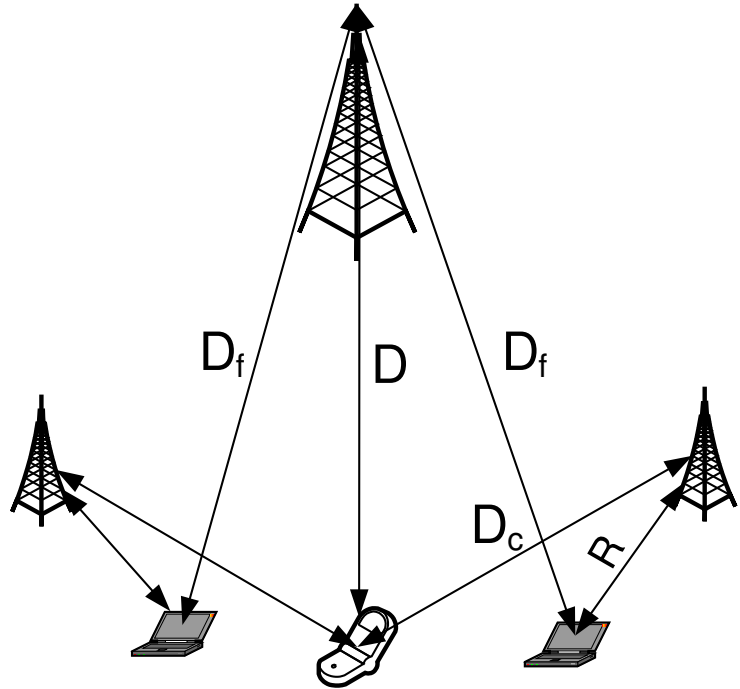}
   \caption{Simple example with $N = 2$ femtocells for determining how link budgets vary with the normalized interference distance $D_f D_c/R_f D$.}
   \label{fig:Example_LinkBudget}
   \end{center}
\end{figure}

\begin{example}[\textbf{$N$ Femtocells}]
Assume a path loss based model wherein the channel gains $g_{i,j} = D_{i,j}^{-\alpha}$ ($D_{i,j}$ represents the distance between user $j$ to BS $B_i$. The term $\alpha$ is the path loss exponent (assumed equal indoors and outdoors for convenience). Femtocell user $i$ is located at distances $R_f$ from its AP $B_i$ and $D_f$ from $B_0$. The cellular user is located at distances $D$ from its macrocell BS $B_0$ and $D_c$ from each femtocell AP (See Fig. \ref{fig:Example_LinkBudget} for $N=2$ femtocells).

In this setup, $\mathbf{q}_c^{T} = \left \lbrack \left(\frac{D_f}{D}\right)^{-\alpha} ,\left(\frac{D_f}{D}\right)^{-\alpha}, \dots, \left(\frac{D_f}{D}\right)^{-\alpha} \right \rbrack$. The vector $\mathbf{q}_f = \left \lbrack \left(\frac{D_c}{R_f}\right)^{-\alpha} ,\left(\frac{D_c}{R_f}\right)^{-\alpha}, \dots, \left(\frac{D_c}{R_f}\right)^{-\alpha} \right \rbrack^{T}$. The decibel link budget $L_{\textrm{dB}}$ varies with $\alpha$ as a straight line and given as
\begin{align}
\label{eq:LBNFemtocells}
L \triangleq \frac{1}{\mathbf{q}_c^{T} \mathbf{q}_F} = \frac{1}{N} \left(\frac{D_f D_c}{D R_f}\right)^{\alpha} \Rightarrow L_{\textrm{dB}} = \underbrace{-10 \log_{10}N}_{\textrm{intercept}}+\underbrace{10\log_{10}\left(\frac{D_f D_c} {D R_f}\right)}_{\textrm{slope}} \alpha.
\end{align}

Define $Q \triangleq \frac{D_f D_c}{D R_f}$ as the \emph{interference distance product normalized by the signaling distance product}. Then, $L_{\textrm{dB}}$ monotonically increases with $\alpha$ whenever the slope $Q_{\textrm{dB}}>0$ and decreases otherwise. Consequently, the condition $Q \gtrless 1$ determines the sensitivity of link budgets to the path-loss exponent.
\end{example}

\subsection{Design Interpretations}
\label{Sec:DesignIntrp}
\begin{figure} [htp]
\begin{center}
   \includegraphics[width=4.0in]{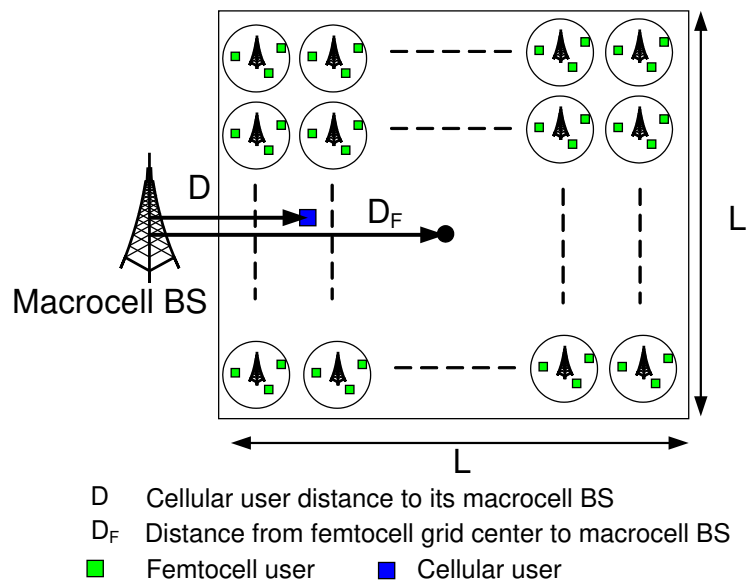}
   \caption{Single transmitting cellular user transmitting in same spectrum with an underlaid grid of femtocells.}
   \label{fig:TwoTier_FemtocellNetwork}
   \end{center}
\end{figure}

This subsection studies how the per-tier SINRs and link budgets vary with user and femtocell locations in practical path loss scenarios. Assume that the cellular user $0$ is located at a distance $D_{0,0}=D$ from the macrocell $B_0$. At a distance $D_{f}$ from $B_0$ (see Fig. \ref{fig:TwoTier_FemtocellNetwork}), $N$ surrounding cochannel femtocells $\{B_i\},i=1 \cdots N$ are arranged in a square grid -- e.g. residential neighborhood -- of area $D_{\textrm{grid}}^2 = 0.25 \textrm{ sq. km.}$ with $\sqrt{N}$ femtocells per dimension. Each femtocell has a radio range equaling $R_f$ meters. Let $D_{i,j}$ denote the distance between transmitting mobile $j$ and BS $B_i$.

For simplicity, neither Rayleigh fading nor lognormal shadowing are modeled.
Assuming a reference distance $D_{\textrm{ref}}=1 \textrm{ meter}$ \cite{Goldsmith} for all users, the channel gains $g_{i,j}$ are represented using the simplified path loss model in the IMT-2000 specification \cite{IMT2000}, given as
\begin{align}
\label{eq:Twotier_Channelgains}
g_{i,j}=\begin{cases}
            K_c \min{(D^{-\alpha_c},1)} & i=j=0, \\
            K_{fi} R_f^{-\beta} & i=j>1, \\
            K_{fo} \phi \min{(D_{0,j}^{-\alpha_{fo}},1)} & i=0, j>0, \\
            K_c \phi \min{(D_{i,j}^{-\alpha_{c}},1)} & i>0, j=0, \\
            K_{fo} \phi^2 \min{(D_{i,j}^{-\alpha_{fo}},1)} &  i \neq j, i,j>0
        \end{cases}
\end{align}
In \eqref{eq:Twotier_Channelgains}, $\alpha_c, \beta, \alpha_{fo}$ respectively denote the cellular, indoor and indoor to outdoor femtocell path loss exponents. Defining $f_{c, \textrm{MHz}}$ as the carrier frequency in MHz, $K_{c,\textrm{dB}} = 30\log_{10}(f_{c,\textrm{Mhz}})-71 $ dB equals the fixed decibel propagation loss during cellular transmissions to $B_0$. The term $K_{fi}$ is the fixed loss between femtocell user $i$ to their BS $B_i$. Finally, $K_{fo}$ denotes the fixed loss between femtocell user $i$ to a different BS $B_j$, and assumed equal to $K_c$. The term $W$ explicitly models partition loss during indoor-to-outdoor propagation (see numerical values for all system parameters in Table \ref{Tbl:SysPrms}).

\begin{assumption}
\label{AS:asPLExponent}
Assume equal outdoor path loss exponents from a cellular user and a femtocell user to the macrocell $B_0$. That is, $\alpha_c = \alpha_{fo}=\alpha$.
\end{assumption}

Following AS\ref{AS:asPLExponent}, substituting \eqref{eq:Twotier_Channelgains} in \eqref{eq:ParetoSINRContourUB2} and assuming that users are at least $1$ meter away from BSs (or $D_{i,j}^{-\alpha}<1 \forall i,j$), the link budget $L$ is given as
\begin{align}
\label{eq:linkbudget}
L = \frac{K_{fi} R_f^{-\beta}}{W^2 K_{fo}} {D^{-\alpha}} \left(\sum \limits_{i=1}^{N} D_{0,i}^{-\alpha}D_{i,0}^{-\alpha}\right)^{-1}.
\end{align}

\begin{figure}[htp]
  \begin{center}
    \includegraphics[width=5in]{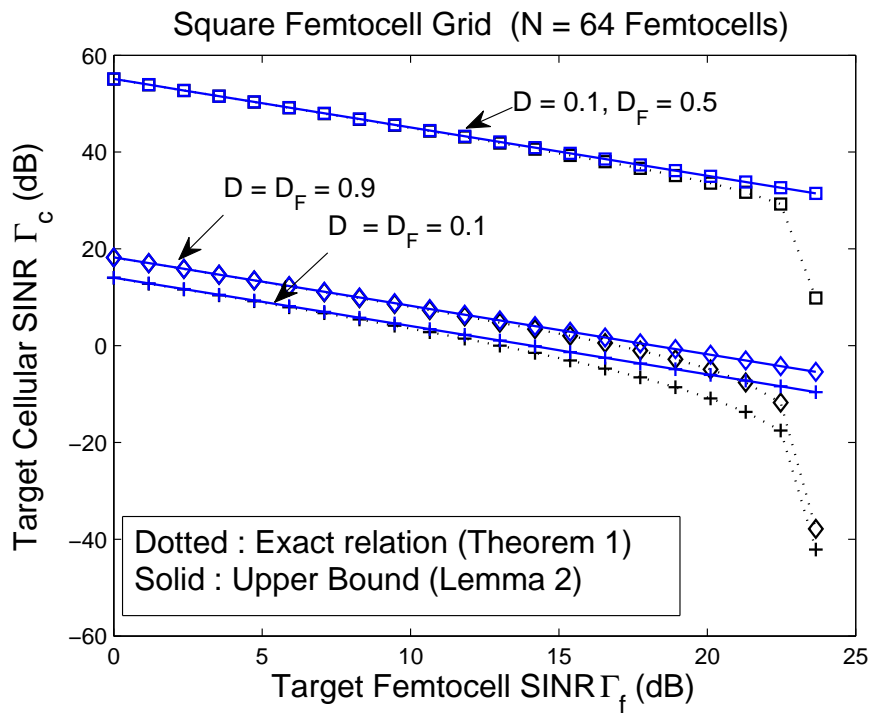}
   \caption{Per-Tier SINR contours for different cellular user and femtocell locations.}
   \label{fig:SINRContoursSquareGrid}
   \end{center}
\end{figure}

Fig. \ref{fig:SINRContoursSquareGrid} shows the SINR contours using \eqref{eq:MacroSINR}, considering a common femtocell SINR target and different normalized $D$ and $D_f$ values. The target spectral radius $\kappa=\rho(\mathbf{\Gamma G})$ was chosen equal to $\max\lbrace 1-10^{-4},\rho(\mathbf{F})+(1-10^{-4})(1-\rho(\mathbf{F})) \rbrace$ (ensuring that $\rho(\mathbf{\Gamma}_f \mathbf{F})<\rho(\mathbf{\Gamma}\mathbf{G}) <1$). For comparison, the upper bound in \eqref{eq:ParetoSINRContourUB2} was also plotted. Three different positions -- normalized w.r.t $R_c$ -- of the cellular user and the femtocell grid are considered namely \begin{inparaenum} [a)]
\item $D=D_{F}=0.1$, \item $D=0.1$ and $D_{F}=0.5$ and
\item $D=D_{F}=0.9$ \end{inparaenum}. In case (a), note that the macrocell BS is located in the \emph{interior} of the femtocell grid.

We observe that employing \eqref{eq:ParetoSINRContourUB2} is a good approximation for the exact result given in \eqref{eq:ParetoSINRContour}. The highest per-tier SINRs occurs in configuration (b) suggesting a low level of normalized interference ($\mathbf{q}_c$ and $\mathbf{q}_f$). Interestingly, when both users and hotspots are close to the macrocell BS [configuration (a)], the per-tier SINRs are \emph{worse} compared to the cell-edge configuration (c). This counterintuitive result suggests that unlike a conventional cellular system where the regular placement of BSs causes the worst-case SINRs typically at cell-edge, the \emph{asymmetric locations of interfering transmissions in a two-tier network potentially diminishes link budgets in the cell-interior as well.} The reason is because power control ``warfare" due to cross-tier interference from femtocells near the macrocell BS necessitates both tiers to lower their SINR targets.

Assuming $D= D_f$ in Fig. \ref{fig:TwoTier_FemtocellNetwork}, the following lemma provides a necessary condition under which the link budget in \eqref{eq:linkbudget} increases with $\alpha$.
\begin{proposition}
\emph{Under assumption \ref{AS:asPLExponent} and assuming fixed locations of all users w.r.t their BSs, the link budget monotonically increases with $\alpha$ whenever
\begin{align}
\label{eq:LinkBudgetMtcty}
\frac{\sum_{i=1}^{N} (D_{0,i}D_{i,0})^{-\alpha} \ln(D_{0,i}D_{i,0})}{\sum_{i=1}^{N}(D_{0,i}D_{i,0})^{-\alpha}} > \ln(D).
\end{align}
}
\end{proposition}
\vspace{2mm}
\begin{proof}
Taking the first derivative of the link budget in \eqref{eq:linkbudget} with respect to $\alpha$ yields \eqref{eq:LinkBudgetMtcty}.
\end{proof}

\begin{figure}[htp]
  \begin{center}
    \subfigure[Link budget for a square femtocell grid configuration.]{\label{fig:LBSquareGrid}    \includegraphics[width=5.0 in]{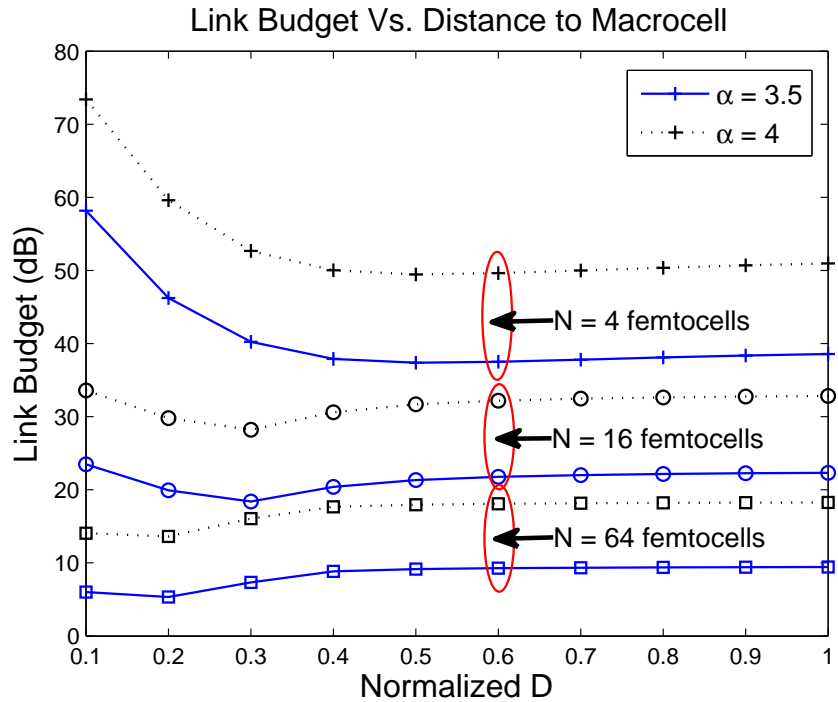}}
    \subfigure[Cumulative distribution function of the link budget $10\log_{10}(1/\mathbf{q}_c^{T}\mathbf{q}_f)$ with randomly located femtocells.]{\label{fig:LBCDFRandom}    \includegraphics[width=5.0 in]{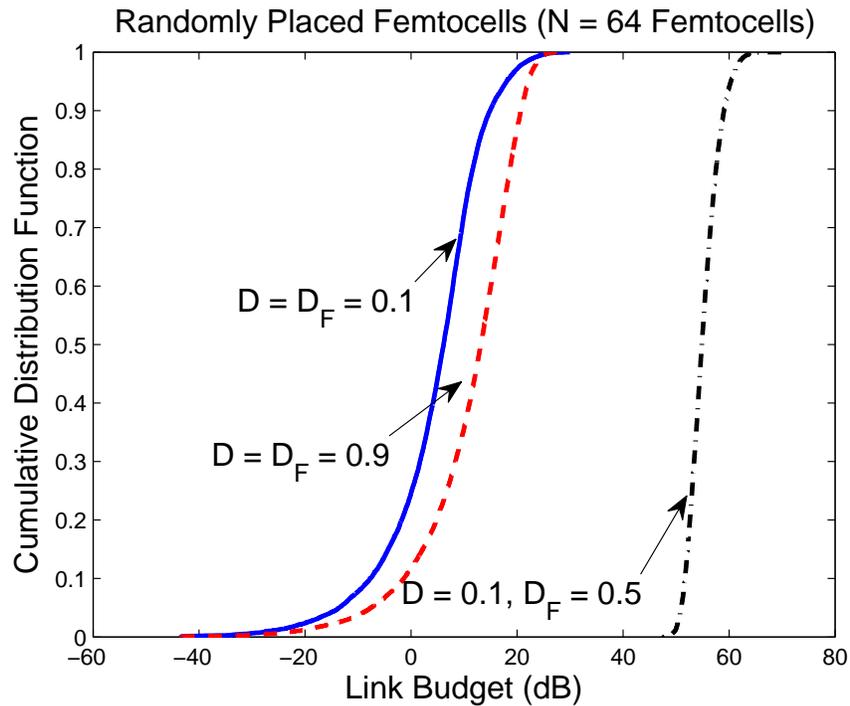}}    \end{center}
  \label{fig:LB}
  \caption{Link budget $10\log_{10}(1/\mathbf{q}_c^{T}\mathbf{q}_f)$ considering a square femtocell grid and randomly placed femtocells.}
\end{figure}

Fig. \ref{fig:LBSquareGrid} plots the Link Budget in \eqref{eq:linkbudget} for $\alpha=3.5,4$ and $N=4,16,64$ femtocells with the cellular user colocated at the grid center ($D = D_F$). The link budgets with $\alpha=4$ are higher relative to those obtained when $\alpha=3.5$ indicating link budgets tend to increase with higher path loss exponents in practical scenarios. Fig. \ref{fig:LBCDFRandom} plots the cumulative distribution function (CDF) of $L_{\textrm{dB}}$ considering randomly distributed femtocells inside a circular region of radius $D_{\textrm{grid}}/\sqrt{\pi}$ centered at distance $D_f$ from $B_0$. With $N = 64$ femtocells, both the regular and random configurations in Figs. \ref{fig:LBSquareGrid}-\ref{fig:LBCDFRandom} show diminishing $L$ in the cell-interior suggesting significant levels of cross-tier interference.

The above results motivate adapting femtocell SINRs with the following objectives namely \begin{inparaenum} \item to maximize their own SINRs, and  \item limit their cross-tier interference. \end{inparaenum}

\section{Utility-Based Distributed SINR Adaptation}
Due to the absence of coordination between tiers, implementing centralized power control $\mathbf{p}^{\ast}=(\mathbf{I}-\mathbf{\Gamma}^{\ast}\mathbf{G})^{-1}\boldsymbol{\eta}^{\ast}$ will likely be prohibitively difficult. In this section, we present a utility-based SINR adaptation scheme.
Using microeconomic concepts, we shall assume that cellular and femtocell users participate in a $N+1$ player non-cooperative power control game $G = [\mathcal{N},\lbrace P_i\rbrace, \lbrace U_i(.) \rbrace]$. Here $\mathcal{N} = \lbrace 0,1,\dots N \rbrace$ refers to the player index set and $P_i$ is the strategy set describing the domain of transmission powers for user $i$. User $i$ maximizes its individual utility $U_i$ (or payoff) in a distributed fashion. Consequently, their actions -- selecting their transmission power -- are the best response to the actions of other participants. For notational convenience, define $[x]^{+} \triangleq \max \lbrace x,0 \rbrace$. Given user $i$, designate $\mathbf{p}_{-i}$ as the vector of transmit powers of all users other than $i$ and define $I_{i}(\mathbf{p}_{-i}) \triangleq \sum_{j \neq i} p_j g_{i,j}+\sigma^2$ as the interference power experienced at $B_i$.

Formally, for all users $0 \leq i \leq N$, this power control game is expressed as
\begin{align}
\label{eq:UtilAdapt}
\max_{0 \leq p_i \leq p_{\textrm{max}}} U_i(p_i,\gamma_i|\mathbf{p}_{-i}) \textrm{ for each user in } \mathcal{N}.
\end{align}
We are interested in computing the equilibrium point (a vector of $N+1$ transmit powers) wherein each user in $\mathcal{N}$ individually maximizes its utility in \eqref{eq:UtilAdapt}, \emph{given} the transmit powers of other users. Such an equilibrium operating point(s) in optimization problem \eqref{eq:FemtoUtilAdapt} is denoted as the \emph{Nash equilibrium} \cite{Nash1951}. Denote $\mathbf{p}^{\ast}=(p_0^{\ast},p_1^{\ast},\dots,p_N^{\ast})$ as the transmission powers of all users under the Nash equilibrium. At the Nash equilibrium, no user can unilaterally improve its individual utility. Mathematically,
\begin{align}
U_i(p_i^{\ast}, \gamma_i^{\ast} | \mathbf{p}_{-i}^{\ast}) \geq U_i(p_i, \gamma_i^{\ast} | \mathbf{p}_{-i}^{\ast}) \quad \forall p_i \neq p_i^{\ast}, p_i \in P_i, \ \forall i \in \mathcal{N}.
\end{align}

We shall make the following assumptions for the rest of the work.
\begin{assumption}
\label{AS:as4}
All mobiles have a maximum transmission power constraint $p_{\textrm{max}}$, consequently the strategy set for user $i$ is given as $P_i = [0,p_{\textrm{max}}]$.
\end{assumption}
\begin{assumption}
\label{AS:as1}
Assume a \emph{closed-loop feedback power control}, i.e BS $B_i$ periodically provides status feedback to user $i \in \mathcal{N}$ if its current SINR $\gamma_i = p_i g_{ii}/I_i(\mathbf{p}_{-i})$ is above/below its minimum SINR target $\Gamma_i$.
\end{assumption}

\subsection{Cellular Utility Function}
Given a current cellular SINR $\gamma_0$ and a minimum SINR target $\Gamma_0 > 0$ at $B_0$, we model the cellular user $0$'s objective as
\begin{align}
\label{eq:MacroUtility}
\max \limits_{0 \leq p_0 \leq p_{\textrm{max}}}{U_0(p_0,\gamma_0|\mathbf{p}_{-0}) = -(\gamma_0-\Gamma_0)^2} .
\end{align}
The intuition behind the strictly concave utility in \eqref{eq:MacroUtility} is that user $0$
desires to achieve its minimum SINR target $\Gamma_0$ -- assuming feasibility -- while expending no more than the minimum required transmission power below $p_\textrm{max}$. Alternatively, given a cellular SINR $\gamma_0 > \Gamma_0$ for a given interference $I_0(\mathbf{p}_{-0})$ at $B_0$, user $0$ could improve its utility by decreasing $p_0$ until $\gamma_0 = \Gamma_0$.

\subsection{Femtocell Utility Function}
Given interfering powers $\mathbf{p}_{-i}$ and current SINR $\gamma_i$, user $i$ in femtocell $B_i$ obtains an individual utility $U_i(p_i,\gamma_i| \mathbf{p}_{-i})$. Having installed the femtocell AP $B_i$ in their self-interest, user $i$ seeks to maximize its individual SINR while meeting its minimum SINR requirement. At the same time, transmitting with too much power will create unacceptable cross-tier interference at the primary infrastructure $B_0$. Consequently, it is natural to discourage femtocells from creating large cross-tier interference. We therefore model the utility function for femtocell user $i$ as consisting of two parts.
\begin{align}
\label{eq:FemtoUtility}
U_i(p_i,\gamma_i| \mathbf{p}_{-i}) = R(\gamma_i,\Gamma_i) + b_i \frac{C(p_i,\mathbf{p}_{-i})}{I_{i}(
\mathbf{p}_{-i})}.
\end{align}
\begin{asparadesc}
\item[Reward function.] The \emph{reward function} $R(\gamma_i,\Gamma_i)$ denotes the payoff to user $i$ as a function of its individual SINR $\gamma_i$ and minimum SINR target $\Gamma_i \leq \frac{p_{\textrm{max}}g_{i,i}}{\sigma^2}$.
\item[Penalty function.] The \emph{penalty function} $b_i \frac{C(p_i,\mathbf{p}_{-i})}{I_{i}(
\mathbf{p}_{-i})}$ is related to the interference experienced at the macrocell BS $B_0$. The penalty $C$ reduces the net utility obtained by $i$ for creating cross-tier interference at $B_0$ by virtue of transmitting at power $p_i$. Here $b_i$ is a constant which reflects the relative importance of the penalty w.r.t the reward of user $i$. Scaling the penalty by $I_i(\mathbf{p}_{-i})$ ensures that femtocells experiencing higher interference are penalized less.
\end{asparadesc}


Using the framework of \cite{JiHuang1998}, we make the following assumptions for femtocell user $i \in \mathcal{N} \setminus \lbrace 0 \rbrace$.
\begin{assumption}
\label{AS:as5}
For the $i$th user, given fixed $p_i$, its utility $U_i(p_i, \gamma_i | \mathbf{p}_{-i})$ is a \emph{monotonically increasing concave upward} function of its SINR $\gamma_i$.
\end{assumption}
\begin{assumption}
\label{AS:as6}
For the $i$th user, given fixed $\gamma_i$, the utility $U_i(p_i, \gamma_i | \mathbf{p}_{-i})$ is a \emph{monotonically decreasing concave downward} function of its transmit power $p_i$.
\end{assumption}
Assumption \ref{AS:as5} models declining satisfaction (marginal utility) obtained by user $i$, once its current SINR $\gamma_i$ exceeds $\Gamma_i$. Assumption \ref{AS:as6} models increased penalty incurred by user $i$ for causing more interference. Under assumptions \ref{AS:as5} and \ref{AS:as6}:
\begin{align}
\label{eq:HongJi1}
\frac{\partial U_i}{\partial \gamma_i}>0  & \Rightarrow \frac{\mathrm{d} R}{\mathrm{d} \gamma_i}>0   &&     \frac{\partial U_i}{\partial p_i}<0 \Rightarrow \frac{\mathrm{d} C}{\mathrm{d} p_i}<0. \\
\label{eq:HongJi2}
\frac{\partial ^2 U_i}{\partial \gamma_i^2}<0 & \Rightarrow \frac{\mathrm{d}^2 R}{\mathrm{d} \gamma_i^2}<0 && \frac{\partial^2 U_i}{\partial p_i^2}<0   \Rightarrow   \frac{\mathrm{d}^2 C}{\mathrm{d} p_i^2} \leq 0.
\end{align}
Taking the second-order total derivative of $U_i$ w.r.t $p_i$ and applying \eqref{eq:HongJi2},
\begin{align}
\label{eq:Concutility}
\frac{\mathrm{d}^2 U_i}{\mathrm{d}p_i^2} = \frac{\mathrm{d} ^2 R}{\mathrm{d} \gamma_i^2}\left(\frac{g_{ii}}{I_i(\mathbf{p}_{-i})}\right)^2+\frac{b_i}{I_i(\mathbf{p}_{-i})}\frac{\mathrm{d}^2 C}{\mathrm{d} p_i^2} <0.
\end{align}
This suggests that given interferer powers $\mathbf{p}_{-i}$, the femtocell utility function $U_i$ at $B_i$ is \emph{strictly concave} with respect to the user $i$'s transmission power $p_i$.

Assume that each femtocell individually maximizes its utility $U(p_i,\gamma_i| \mathbf{p}_{-i})$ as a best response to the cellular user and neighboring femtocell users' transmit powers $\mathbf{p}_{-i}$. The problem statement is given as
\begin{align}
\label{eq:FemtoUtilAdapt}
\max_{0 \leq p_i \leq p_{\textrm{max}}} U_i(p_i, \gamma_i | \mathbf{p}_{-i}) =
\max_{0 \leq p_i \leq p_{\textrm{max}}} \left \lbrack R(\gamma_i,\Gamma_i) + b_i \frac{C(p_i,\mathbf{p}_{-i})}{I_{i}(\mathbf{p}_{-i})} \right \rbrack.
\end{align}

\subsection{Existence of Nash Equilibrium}
Observe that for all $i \in \mathcal{N}$, $U_i$ is continuous in $\mathbf{p}$ and $U_i$ is strictly concave w.r.t $p_i$ from \eqref{eq:Concutility} over a convex, compact set $[0, p_{\textrm{max}}]$. We now employ the following theorem from Glicksberg \cite{Glicksberg52}, Rosen \cite{Rosen65} and Debreu \cite{Debreu52}:
\begin{theorem}
\label{Th:ExistenceNE}
\emph{A Nash equilibrium exists in game $G = [\mathcal{N},\lbrace P_i \rbrace,\lbrace U_i(.)\rbrace]$ if, for all $i = 0,1,\dots,N$,
\begin{enumerate}
\item $P_i$ is a nonempty, convex and compact subset of some Euclidean space $\mathbb{R}^{N+1}$.
\item $U_i(\mathbf{p})$ is continuous in $\mathbf{p}$ and quasi-concave in $p_i$.
\end{enumerate}}
\end{theorem}

Following Theorem \ref{Th:ExistenceNE}, the optimization problems in \eqref{eq:MacroUtility} and \eqref{eq:FemtoUtilAdapt} have a Nash Equilibrium. The following theorem derives the SINR equilibria at each femtocell.
\begin{theorem}
\label{Th:FemtoNashEqb}
\emph{
A SINR Nash equilibrium at femtocell BS $B_i, i \in \mathcal{N} \setminus \lbrace 0 \rbrace$ satisfies $\gamma_i^{\ast} = p_i^{\ast}g_{i,i}/I_i(\mathbf{p}_{-i}^{\ast})$, where $p_{i}^{\ast}$ is given as
\begin{align}
\label{eq:NEGeneralFemtoUtility}
p_i^{\ast}= \min{\left\lbrace \left \lbrack \frac{I_i(\mathbf{p}_{-i}^{\ast})}{g_{i,i}}f_i^{-1}\left(-\frac{b_i}{g_{i,i}}\frac{\mathrm{d}C}{\mathrm{d}p_i}\right) \right \rbrack^{+},p_{\textrm{max}} \right \rbrace} \textrm{ and } f_i(x) \triangleq \left \lbrack \frac{\mathrm{d} R(\gamma_i,\Gamma_i)}{\mathrm{d}\gamma_i} \right \rbrack_{\gamma_i = x}.
\end{align}
}
\end{theorem}
\vspace{2mm}
\begin{proof}
Since femtocell user $i$ individually optimizes its utility as a best response to other users, we first fix interfering powers $\mathbf{p}_{-i}$. Because $U_i(p_i,\gamma_i|\mathbf{p}_{-i})$ is a strictly concave function of $p_i$, its partial derivative $U_i'(p_i,\gamma_i|\mathbf{p}_{-i})$ -- assuming differentiability -- monotonically decreases with increasing $p_i$. A necessary condition for the existence of local optima is that the derivative of $U_i$ in the interval $[0,p_{max}]$ equals zero. Therefore, if there is no local optima in the interval $[0,p_{\textrm{max}}]$, the user $i$ chooses its equilibrium transmit power $p_i^{\ast}$ depending on the sign of the derivative $U_i'(p_i,\gamma_i)$ -- transmit at full power (if $U_i'(p_i,\gamma_i) > 0 $ in $[0,p_{\textrm{max}}]$) or zero power otherwise.

On the contrary, if the Nash equilibrium ${p}_i^{\ast}$ is a local optima in $[0,p_{\textrm{max}}]$,
\begin{align}
\label{eq:NENeccCond}
\left \lbrack \frac{\mathrm{d}U_i(p_i, \gamma_i | \mathbf{p}_{-i})}{\mathrm{d}p_i}\right \rbrack_{p_i = p_i^{\ast}} =0 \Rightarrow
\left \lbrack \frac{\mathrm{d} R(\gamma_i,\Gamma_i)}{\mathrm{d} \gamma_i}\frac{g_{i,i}}{I_i(\mathbf{p}_{-i})}+
\frac{b_i}{I_{i}(\mathbf{p}_{-i})}\frac{\mathrm{d}C}{\mathrm{d}p_i} \right \rbrack_{p_i = p_i^{\ast}} =0 \ \forall i \in \mathcal{N}, i \geq 1.
\end{align}
Since $I_i(\mathbf{p}_{-i}) \geq \sigma^2 >0$, one may cancel $I_i(\mathbf{p}_{-i})$ on both sides of \eqref{eq:NENeccCond}. The conditions \eqref{eq:HongJi1}-\eqref{eq:HongJi2} ensure that $\mathrm{d}R(\gamma_i,\Gamma_i)/\mathrm{d}\gamma_i$ [resp. $-\mathrm{d}C/\mathrm{d}p_i$] are monotone decreasing [resp. monotone non-decreasing] in $p_i$. The solution to \eqref{eq:NENeccCond} corresponds to the intersection of a monotone decreasing function $g_{i,i}\mathrm{d}R(\gamma_i,\Gamma_i)/\mathrm{d}\gamma_i$ and a monotone increasing function $-b_i\mathrm{d}C/\mathrm{d}p_i$ w.r.t the transmitter power $p_i$. Given $\mathbf{p}_{-i}^{\ast}$, this intersection is unique \cite[Section 3]{JiHuang1998} and corresponds to the Nash equilibrium at $p_i=p_i^{\ast}$. Using the notation $f_i(x) \triangleq \left \lbrack \frac{\mathrm{d} R(\gamma_i,\Gamma_i)}{\mathrm{d}\gamma_i} \right \rbrack$ evaluated at $\gamma_i = x$ yields \eqref{eq:NEGeneralFemtoUtility}. This completes the proof.
\end{proof}

\subsubsection{Femtocell Utility Selection}
Assume the $R(\gamma_i,\Gamma_i)$ and $C(p_i,\mathbf{p}_{-i})$ in \eqref{eq:FemtoUtility} as shown below.
\begin{align}
\label{eq:ExpoLinearFemtoUtility}
R(\gamma_i,\Gamma_i) = 1-e^{-a_i(\gamma_i-\Gamma_i)}, \gamma_i \geq 0,\
C(p_i,\mathbf{p}_{-i}) = -p_i g_{0,i}.
\end{align}
The exponential reward intuitively models femtocell users' desire for higher SINRs relative to their minimum SINR target. The linear cost $C(p_i,\mathbf{p}_{-i}) = -p_i g_{0i}$  discourages femtocell user $i$ from decreasing the cellular SINR by transmitting at high power. Assuming $a_i,b_i \neq 0$, it can be verified that the above choice of $R(\gamma_i,\Gamma_i)$ and $C(p_i,\mathbf{p}_{-i})$ satisfies the conditions outlined in \eqref{eq:HongJi1} and \eqref{eq:HongJi2}.
\begin{align}
\frac{\mathrm{d} R}{\mathrm{d} \gamma_i}=a_i e^{-a_i(\gamma_i-\Gamma_i)} >0   &&     \frac{b_i}{I_i(\mathbf{p}_{-i})}\frac{\mathrm{d} C}{\mathrm{d} p_i}=-\frac{b_i g_{0,i}}{I_i(\mathbf{p}_{-i})}<0\\
\frac{\mathrm{d}^2 R}{\mathrm{d} \gamma_i^2}=-a_i^2 e^{-a_i(\gamma_i-\Gamma_i)} <0 && \frac{b_i}{I_i(\mathbf{p}_{-i})}\frac{\mathrm{d}^2 C}{\mathrm{d} p_i^2} = 0.
\end{align}

\begin{lemma}
\emph{With the utility-based cellular SINR adaptation [resp. femtocell SINR adaptation] in \eqref{eq:MacroUtility} [resp. \eqref{eq:FemtoUtilAdapt} with reward-cost functions in \eqref{eq:ExpoLinearFemtoUtility}], the unique SINR equilibria at BS $B_i, i \in \mathcal{N}$ are given as $\gamma_i^{\ast} = \frac{p_i^{\ast}g_{ii}}{I_i(\mathbf{p}_{-i})}$ where $p_i^{\ast}$ is given as
\begin{align}
\label{eq:NEExpoLinearFemtoUtility}
\textrm{Femtocell User : }  p_i^{\ast} &= \min \left \lbrace \frac{I_i(\mathbf{p}_{-i}^{\ast})}{g_{i,i}} \left \lbrack \Gamma_i + \frac{1}{a_i}\ln{\left(\frac{a_i g_{i,i}}{b_i g_{0,i}} \right)} \right \rbrack^{+},p_{\textrm{max}} \right \rbrace. \\
\label{eq:NECellular}
\textrm{Cellular User : }  p_0^{\ast} &= \min \left \lbrace \frac{I_0(\mathbf{p}_{-0}^{\ast})}{g_{0,0}}  \Gamma_0,p_{\textrm{max}} \right \rbrace.
\end{align}
}
\end{lemma}
\vspace{2mm}
\begin{proof}
The cellular user's utility function $U_0(p_0,\gamma_0|\mathbf{p}_{-0})$ is strictly concave w.r.t $p_0$ given $\mathbf{p}_{-0}$. Consequently, the argument maximizer in \eqref{eq:MacroUtility} occurs either in the interior at $p_0^{\ast} = \Gamma_0 \frac{I_0(\mathbf{p}_{-0})}{g_{00}}$ or at the boundary point $p= p_{\textrm{max}}$ if $U_0'(p_0,\gamma_0|\mathbf{p}_{-0})=2\frac{g_{00}}{I_0(\mathbf{p}_{-0}^{\ast})}(\Gamma_0-p_0\frac{g_{00}}{I_0(\mathbf{p}_{-0}^{\ast})})>0$ in $[0,p_\textrm{max}]$. At femtocell AP $B_i$, the equilibrium SINR in Equation \eqref{eq:NEExpoLinearFemtoUtility} follows immediately by applying \eqref{eq:NEGeneralFemtoUtility} in Theorem \ref{Th:FemtoNashEqb} to the utility functions given in \eqref{eq:ExpoLinearFemtoUtility}.

To show uniqueness of the Nash equilibria, we rewrite Equations \eqref{eq:NEExpoLinearFemtoUtility}-\eqref{eq:NECellular} as an iterative power control update $\mathbf{p}^{(k+1)}=\mathbf{f}(\mathbf{p}^{(k)})$ -- wherein the component $f_i(p_i)$ represents the power update for user $i$ -- with individual power updates given as
\begin{align}
\label{eq:SINRFemtoIterative}
\textrm{Femtocell User : } p_i^{(k+1)}&=\min\left\lbrace\frac{p_i^{(k)}}{\gamma_i^{(k)}} \left \lbrack \Gamma_i + \frac{1}{a_i}\ln{\left(\frac{a_i g_{i,i}}{b_i g_{0,i}} \right)} \right \rbrack^{+},p_{\textrm{max}} \right \rbrace. \\
\label{eq:SINRMacroIterative}
\textrm{Cellular User : } p_0^{(k+1)}&=\min\left\lbrace\frac{p_0^{(k)}}{\gamma_i^{(k)}} \Gamma_0,p_{\textrm{max}} \right \rbrace.
\end{align}

Yates \cite{Yates1995a} has shown that, provided a power control iteration of the form $\mathbf{p}^{(k+1)} = \mathbf{f}(\mathbf{p}^{(k)})$ has a fixed point and whenever $\mathbf{f}(\mathbf{p})$ satisfies the following properties namely \begin{inparaenum}[a)] \item positivity $f(\mathbf{p})>0$, \item monotonicity $\mathbf{p}_1 > \mathbf{p}_2 \Rightarrow \mathbf{f}(\mathbf{p}_1)>\mathbf{f}(\mathbf{p}_2)$ and \item scalability $\alpha \mathbf{f}(\mathbf{p}) > \mathbf{f}(\alpha \mathbf{p}) \ \forall \alpha>1$\end{inparaenum}, then the power control iteration converges to the fixed point, which is unique. In such a case, $\mathbf{f}$ is called a \emph{standard interference function}.
Since the RHSs in \eqref{eq:SINRFemtoIterative}-\eqref{eq:SINRMacroIterative} form a standard interference function, its fixed point (or the Nash equilibrium given by \eqref{eq:NEExpoLinearFemtoUtility}-\eqref{eq:NECellular}) is \emph{unique} and the iterates are guaranteed to converge to the equilibrium transmit powers. This completes the proof.
\end{proof}
In a practical tiered cellular deployment, \eqref{eq:SINRFemtoIterative} can be implemented in a distributed fashion since each femtocell user $i$ only needs to know its own target SINR $\Gamma_i$ and its channel gain to $B_0$ and $B_i$ given as $g_{0i}$ and $g_{ii}$ respectively.
Estimating $g_{0,i}$ at femtocell $B_i$ may require site specific knowledge\cite{Chen2007}. Possibly, femtocells would infer their locations using indoor GPS, or even estimate the path losses from the macrocell downlink signal in a TDD system (assuming reciprocity).

\begin{remark}
Given equal minimum SINR targets at all femtocells and assuming identical coefficients in the utility functions ($a_i=a, b_i=b \ \forall i \in \mathcal{N}\setminus \lbrace 0 \rbrace $), femtocell users with higher $g_{i,i}/g_{0,i}$ (or a higher received signal strength relative to cross-tier macrocell interference) obtain a higher relative improvement in their SINR equilibria.
\end{remark}
\vspace{2mm}
The choice of the coefficients $a_i$ and $b_i$ entails careful consideration of the trade-offs between the femtocell users' desire to maximize their own data rates and the relative importance of satisfying the cellular users' QoS requirement. The Nash equilibrium defined in \eqref{eq:NEExpoLinearFemtoUtility} has the following properties.
\begin{enumerate}
\item For large $a_i$ ($a_i \rightarrow \infty$), the equilibria $\gamma_i^{\ast} \rightarrow \Gamma_i$ (assuming $\Gamma_i$ is feasible $\forall i$, that is, \eqref{eq:Twotier_PFcondition} is satisfied). This corresponds to hotspot users with \emph{little inclination} to exceed their minimum rate requirement (e.g. voice users). In such a case, \eqref{eq:SINRFemtoIterative} is equivalent to the Foschini-Miljanic (FM) algorithm $p_i^{(k+1)} = \min \left \lbrace p_i^{(k)}\frac{\Gamma_i}{\gamma_i^{(k)}}, p_{\textrm{max}} \right \rbrace$ \cite{Foschini1993,Grandhi1994}.

\item If $a_i$ is chosen such that $a_i g_{i,i} < b_i g_{0,i}$, the hotspot users' SINR equilibria are lesser than their minimum target $\Gamma_i$, because they pay a greater penalty for causing cross-tier macrocell interference.
\item Choosing $a_i <1$ and $\frac{a_i}{b_i} \gg 1$ increases the importance provided to the reward function relative to the cost function at each femtocell. Indeed, taking the derivative of $\frac{1}{a_i} \ln \left(\frac{a_i g_{i,i}}{b_i g_{0,i}}\right)$ w.r.t $a_i$ yields
\begin{align}
\frac{\textrm{d}}{\textrm{d}a_i} \left \lbrack \frac{1}
{a_i}\ln \left(\frac{a_i g_{i,i}}{b_i g_{0,i}}\right) \right\rbrack=\frac{1}{a_i^2}\left(1-\ln \left(\frac{a_i g_{i,i}}{b_i g_{0,i}}\right) \right) >0 \ \forall \frac{a_i g_{i,i}}{b_i g_{0,i}}<e = 2.71828\dots
\end{align}
Therefore, the highest gains over the minimum SINR target $\Gamma_i$ are obtained when $a_i g_{i,i}=e b_i g_{0,i}$. Such a choice is not necessarily preferable since the potentially large cross-tier interference from femtocells may result in $\gamma_0^{\ast} < \Gamma_0$.
\end{enumerate}

\subsection{Reducing Femtocell SINR Targets : Cellular Link Quality Protection}
Whenever the cellular SINR target $\Gamma_0$ is infeasible, user $0$ transmits with maximum power according to \eqref{eq:SINRMacroIterative}. Assume, after the $M$th iterate (assuming large $M$), user $0$'s SINR $\gamma_0^{(M)}<(1-\epsilon)\Gamma_0$ where $\epsilon$ is a pre-specified SINR tolerance for the cellular user.
\begin{align}
\label{eq:FemtoSINRReduceA}
(1-\epsilon)\Gamma_0 > \gamma_0^{(M)} &= \frac{p_{\textrm{max}}g_{0,0}}{\sum\limits_{i=1}^{N}p_i^{(M)} g_{0,i}+\sigma^2}.
\end{align}

For guaranteeing that user $0$ achieves its SINR target within its tolerance, that is $\gamma_0^{(M)} \geq (1-\epsilon)\Gamma_0$, we propose that a femtocell subset $\Pi \subseteq \lbrace B_1,B_2,\dots,B_N \rbrace$ reduce their SINR equilibria in \eqref{eq:NEExpoLinearFemtoUtility} by a factor $t>1$.  A centralized selection of $t$ ensures
\begin{align}
\label{eq:FemtoSINRReduceB}
(1-\epsilon)\Gamma_0 \leq \frac{p_{\textrm{max}}g_{0,0}}{\frac{1}{t}\sum\limits_{i: B_i \in \Pi}p_i^{(M)} g_{0,i}+\sum\limits_{j: B_j \in \Pi^{C}}p_j^{(M)} g_{0,j}+\sigma^2}
\end{align}
where $\Pi^{C}$ denotes the set complement of $\Pi$. Combining \eqref{eq:FemtoSINRReduceA} \& \eqref{eq:FemtoSINRReduceB}, a sufficient condition to obtain $\gamma_0 \geq \Gamma_0$ at $B_0$ is that there exists $t>1$ and $\Pi \subseteq \lbrace B_1,B_2,\dots,B_N \rbrace$ such that
\begin{align}
\label{MacroSINRSatisfiability}
    \left(1-\frac{1}{t}\right)\sum_{i:B_i \in \Pi}p_i^{(M)}g_{0,i} \geq p_{\textrm{max}}g_{0,0}\left(\frac{1}{\gamma_0^{(M)}}-\frac{1}{(1-\epsilon)\Gamma_0}\right).
\end{align}
In \eqref{MacroSINRSatisfiability}, whenever $\Pi_1 \subseteq \Pi_2 \subseteq \lbrace B_1, \dots B_N \rbrace$, then $t_{\Pi_1} \geq t_{\Pi_2}$. That is, choosing an expanding set of femtocell BSs to reduce their SINR targets requires a monotonically decreasing SINR reduction factor for each femtocell. Further, if reducing SINR targets inside a femtocell set $\Pi_1$ does not achieve $\Gamma_0$ at $B_0$, then a bigger femtocell set $\Pi_2 \supset \Pi_1$ should be chosen. Centralized selection of $t$ and $\Pi$ may be practically hard especially in two-tier networks employing OFDMA because the macrocell BS may need to communicate the $t$'s and $\Pi$ sets for each frequency sub band. A simpler strategy is to distributively adapt the femtocell SINR targets based on periodic feedback from the macrocell BS.
\begin{assumption}
\label{AS:as7}
Following every $M$th update in \eqref{eq:SINRFemtoIterative}, an SINR status feedback occurs from $B_0$ to $B_i$'s whether $\gamma_0^{(M)} < (1-\epsilon)\Gamma_0$.
\end{assumption}
Given $M$ iterative updates, define the set $\Pi_{(M)}$ [resp. its complement $\Pi^{c}_{(M)}$] as the \emph{dominant [resp. non-dominant] interferer} set, consisting of femtocells whose interference at $B_0$ individually exceeds [resp. below] a threshold $y>0$. Mathematically, $\Pi_{(M)}(y) \triangleq \lbrace B_i: p_i^{(M)} g_{0,i} > y \rbrace$. Whenever femtocell user $i$ determines that $B_i \in \Pi(y)$, it scales down its SINR target $\gamma_i^{\ast}$ in \eqref{eq:NEExpoLinearFemtoUtility} by $t>1$. Denoting the set cardinality by $\lvert X \rvert$, the above selection chooses the $\lvert \Pi(y) \rvert$ strongest femtocell interferers for reducing their transmit powers. Periodically decreasing $y$ by a factor $\delta y$ after every $M$ iterations increases $|\Pi(y)|$. Specifically, for all $j \geq i$, choosing $y_{Mj} \leq y_{Mi}$ ensures that $\Pi_{Mj} \supseteq \Pi_{Mi}$. Given a tolerance $\epsilon$, the SINR reduction procedure is repeated after every $M$ updates until the cellular user's SINR is greater than $(1-\epsilon)\Gamma_0$. See Algorithm \ref{Al:MaintainLQMacro} for the pseudocode. Table \ref{Tbl:Example} shows the algorithm performance in a practical scenario of a macrocell overlaid with $16$ femtocells.

Provided the SINR at $B_0$ equals $(1-\epsilon)\Gamma_0$, the \emph{mean femtocell dB SINR $\langle \gamma_{\textrm{dB}}^{\ast} \rangle$}, the \emph{average percentage of degraded femtocells $\langle N \rangle $} and the \emph{average percentage dB SINR degradation $\langle \Delta(\gamma^{\ast}) \rangle$} at femtocells (assuming zero SINR degradation at femtocells with $\gamma_i^{\ast} \geq \Gamma_i$) can be calculated as:
\begin{align}
\label{eq:Perfmetrics}
\langle \gamma_{\textrm{dB}}^{\ast} \rangle &= \frac{1}{N}\sum_{i=1}^{N} 10 \log_{10} \gamma_i^{\ast}. \notag \\
 \langle |\Pi |\rangle &= \frac{1}{N}\left \lvert \lbrace B_i \in \Pi: \gamma_i^{(M)}<\Gamma_i \rbrace \right \rvert. \notag \\
\langle \Delta(\gamma^{\ast}) \rangle &= \left \lbrack\frac{1}{N}\sum_{B_i \in \Pi: \gamma_i^{(M)}<\Gamma_i}\frac{10\log_{10}\Gamma_i-10\log_{10}{\gamma_i^{(M)}}}{10\log_{10}\Gamma_i} \right \rbrack.
\end{align}

\section{Numerical Results}
In this section, we present numerical results based on two experiments with the system parameters in Table \ref{Tbl:SysPrms} and the setup in Section \ref{Sec:DesignIntrp}. The AWGN power $\sigma^2$ in \eqref{eq:Twotier_SINR} was determined after assuming a cell-edge user obtains a cellular SNR equaling $20$ dB at $B_0$, while employing maximum transmission power. Results are reported for $5000$ different SINR trials in each experiment. The minimum femtocell SINR targets were randomly selected (uniform distribution) in the interval $[\Gamma_{f,\textrm{min}},\Gamma_{f,\textrm{max}}]$ dB. In any given trial, if the generated set of minimum SINR targets $\mathbf{\Gamma}_f$ resulted in $\rho(\mathbf{\Gamma}_f \mathbf{F})>1$ in \eqref{eq:GMtxDecomp}, then our experiments scaled $\mathbf{\Gamma}_f$ by a factor $\rho(\mathbf{\Gamma}_f \mathbf{F})(1+10^{-3})$ for ensuring feasible femtocell SINR targets.

The first experiment obtains the improvements in femtocell SINRs relative to their minimum SINR targets with our proposed SINR adaptation. A cell-edge location of the cellular user ($D = 0.9$) and the femtocell grid ($D_F = 0.9$) is considered. To maximize the chance of obtaining a feasible set of $(N+1)$ SINRs, the cellular SINR target $\Gamma_0$ is equal to either its minimum target $\Gamma_{c,\textrm{min}}=3$ dB, or scaling its highest obtainable target in \eqref{eq:MacroSINR} by $\Delta_{c,\textrm{dB}}=  5$ dB (which ever is larger) and given as
\begin{align}
\Gamma_0 = \max \left \lbrace \Gamma_{c,\textrm{min}}, \frac{1}{\Delta_c}\frac{\kappa^2}{\mathbf{q}_c^{T}[\mathbf{I}-(\mathbf{\Gamma}_f/ \kappa)\mathbf{F}]^{-1}\mathbf{\Gamma}_f \mathbf{q}_f} \right \rbrace.
\end{align}

\begin{figure}[htp]
  \begin{center}
    \includegraphics[width=5in]{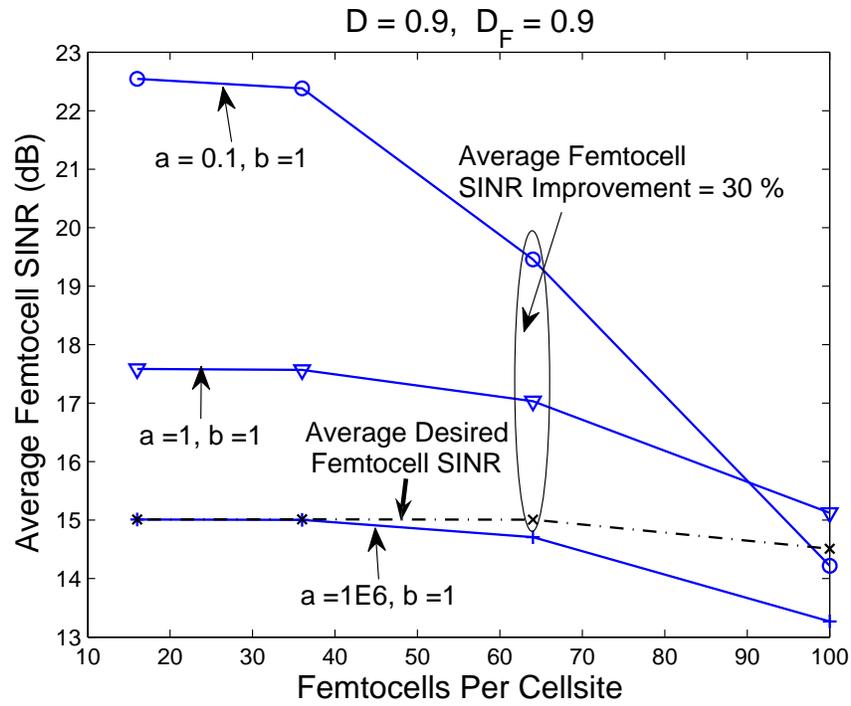}
  \caption{Mean femtocell SINR targets (grid center at cell-edge) for different reward and cost coefficients.}
  \label{fig:MeanFemtoSINR}
  \end{center}
\end{figure}

Assuming $a_i = a$ and $b_i = b \ \forall i \geq 1$ in \eqref{eq:NEExpoLinearFemtoUtility}, Fig. \ref{fig:MeanFemtoSINR} plots the mean decibel femtocell SINRs ($D = D_f = 0.9$) in \eqref{eq:Perfmetrics} for different $a$ and $b$ values. Selecting $a<1$ models femtocell users seeking a greater SINR reward relative to their minimum SINR target. With $a = 0.1, b = 1$ and $N = 64$ femtocells, there is a nearly $30$ \% improvement in mean femtocell SINRs relative to their average minimum SINR target. With a higher interference penalty at femtocells ($b = 1$), our utility adaptation yields a nearly $2$ dB improvement in mean femtocell SINRs above their mean SINR target. When $a >> 1$, femtocell users have little inclination to exceed their minimum SINR targets. In fact, with $N \geq 64$ femtocells, the mean equilibrium femtocell SINRs are \emph{below the mean SINR target} because femtocell users turn down their transmit powers to improve the cellular link quality.

The second experiment considers randomly selected decibel cellular SINR targets chosen uniformly in the interval $[\Gamma_{c,\textrm{min}},\Gamma_{c,\textrm{max}}]$ dB. All femtocells selected identical coefficients $a_i = b_i = 1$ in in \eqref{eq:NEExpoLinearFemtoUtility}. Femtocells scaled down their SINR targets in \eqref{eq:SINRFemtoIterative} until the cellular user $0$ approached within $95$\% of its minimum SINR target.

\begin{figure}[htp]
  \begin{center}
    \includegraphics[width=5in]{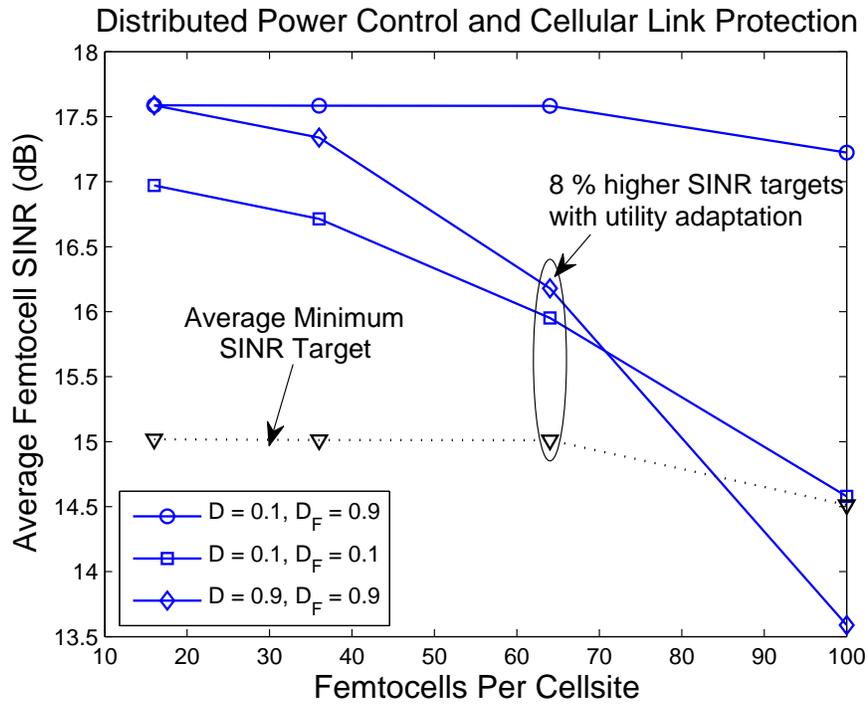}
  \caption{Mean femtocell SINR targets with distributed power control and cellular link quality protection.}
  \label{fig:MeanFemtoSINR_DPC_LinkProt}
  \end{center}
\end{figure}

Figs. \ref{fig:MeanFemtoSINR_DPC_LinkProt} shows the average femtocell decibel SINRs $\langle \gamma_{\textrm{dB}}^{\ast} \rangle $ using the distributed power control in \eqref{eq:SINRFemtoIterative}-\eqref{eq:SINRMacroIterative} and cellular link quality protection. The black dotted lines plot the average minimum femtocell SINR target $10\log_{10}(\sqrt{\Gamma_{f,\textrm{min}}\Gamma_{f,\textrm{max}}})$. Fig. \ref{fig:MeanFemtoSINR_DPC_LinkProt} shows that with $N=64$ femtocells, a nearly $8\%$
 SINR improvement is obtained when the user and femtocells are located on the cell-edge.

\begin{figure}[htp]
  \begin{center}
    \subfigure[Mean percentage SINR reduction at femtocells.]{\label{fig:FemtoSINRDgrd}\includegraphics[width = 5in]{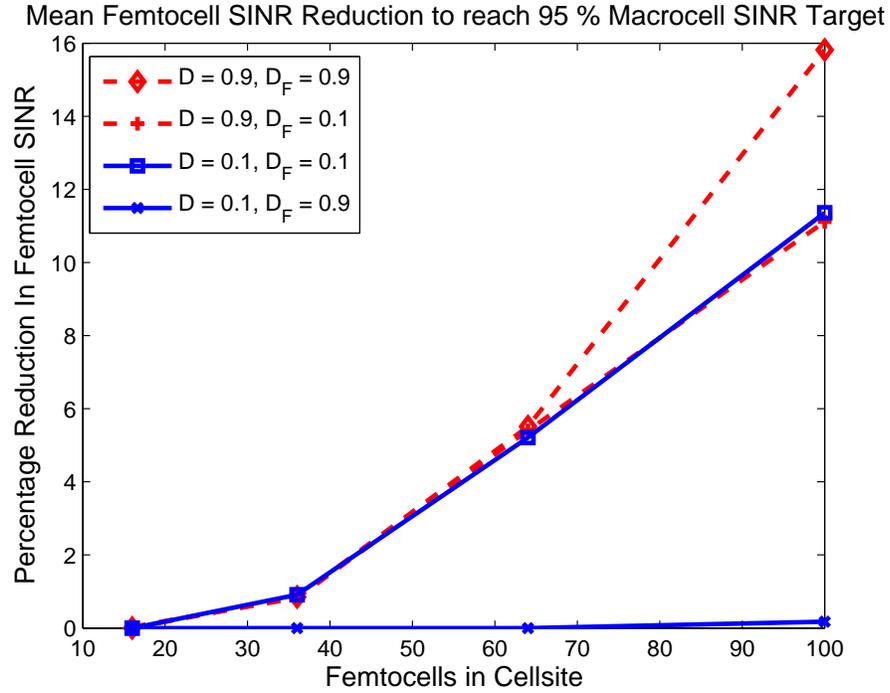}}
    \subfigure[Mean percentage of femtocells below their SINR target.]{\label{fig:NumDgrdFemtos}\includegraphics[width = 5in]{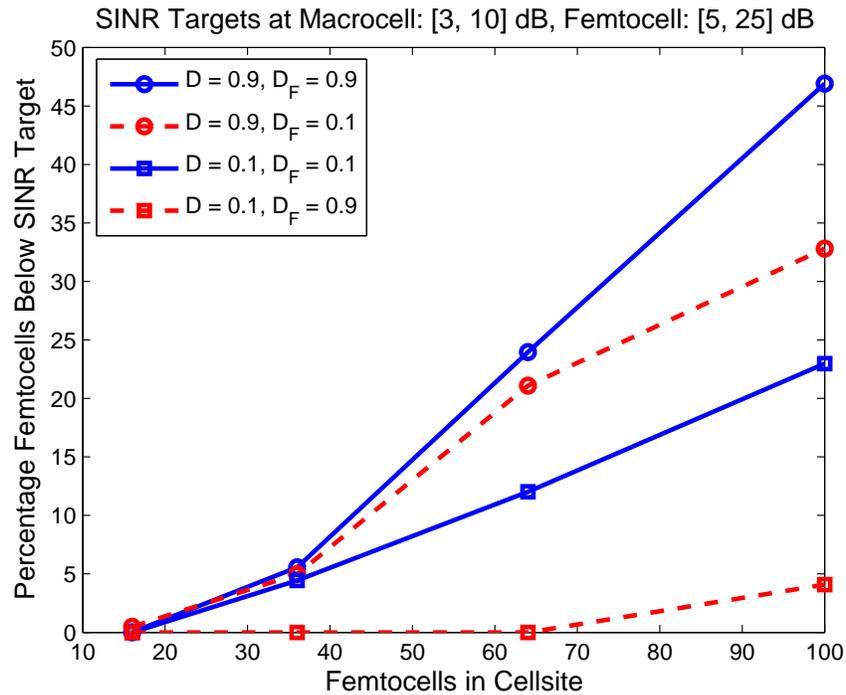}}    \end{center}
  \label{fig:FemtocellSINR}
  \caption{Femtocell SINR reduction when cellular SINR target is uniformly distributed in $[3,10]$ dB, and initial femtocell SINR targets are uniformly distributed in $[5,25]$ dB.}
\end{figure}

Figs. \ref{fig:FemtoSINRDgrd}-\ref{fig:NumDgrdFemtos} plot the mean percentage reduction in femtocell SINRs $ \langle \Delta(\gamma^{\ast}) \rangle$ and the mean percentage of ``degraded" femtocells $ \langle |\Pi | \rangle $ in \eqref{eq:Perfmetrics}. With $N = 100$ femtocells and a cell-edge location ($D = 0.9, D_F =0.9$), although Fig. \ref{fig:NumDgrdFemtos} shows that nearly $45\%$ of femtocells operate below their minimum SINR target, the worst-case femtocell SINR reduction at femtocells is only $16\%$  [Fig. \ref{fig:FemtoSINRDgrd}]. In all other cases, the mean percentage SINR reduction is less than $6\%$. This shows that our cellular link quality protection algorithm guarantees reliable cellular coverage without significantly affecting femtocell SINR targets.

\section{Conclusion}
Cellular operators will obtain better spectral usage and reduced costs by deploying macrocell and femtocell users in a shared region of spectrum. Our work has addressed three related questions. The first is that of determining the radio link quality for a cellular user, given a set of $N$ transmitting femtocells with different SINR targets. The takeaway is that achieving higher SINR targets in one tier fundamentally constricts the highest SINRs obtainable in the other tier. The reason is because of near-far effects caused by the asymmetric positions of interfering users w.r.t nearby BSs. The second and third questions seek to determine femtocell data rates when home users perform utility-based SINR adaptation; providing link quality protection to an active cellular user may necessitate femtocells to deliberately lower their SINR targets. We provide a link quality protection algorithm for progressively reducing the SINR targets at strong femtocell interferers when a cellular user is unable to meet its SINR target. Simulation results confirm the efficacy of the proposed algorithm and its minimal impact (worst case femtocell SINR reduction of only $16 \%$) on femtocell SINRs. Being distributed, the power control algorithm ensures minimal network overhead in a practical two-tier deployment.

\bibliographystyle{IEEEtran}

\begin{thebibliography}{10}
\providecommand{\url}[1]{#1}
\csname url@rmstyle\endcsname
\providecommand{\newblock}{\relax}
\providecommand{\bibinfo}[2]{#2}
\providecommand\BIBentrySTDinterwordspacing{\spaceskip=0pt\relax}
\providecommand\BIBentryALTinterwordstretchfactor{4}
\providecommand\BIBentryALTinterwordspacing{\spaceskip=\fontdimen2\font plus
\BIBentryALTinterwordstretchfactor\fontdimen3\font minus
  \fontdimen4\font\relax}
\providecommand\BIBforeignlanguage[2]{{%
\expandafter\ifx\csname l@#1\endcsname\relax
\typeout{** WARNING: IEEEtran.bst: No hyphenation pattern has been}%
\typeout{** loaded for the language `#1'. Using the pattern for}%
\typeout{** the default language instead.}%
\else
\language=\csname l@#1\endcsname
\fi
#2}}

\bibitem{ChandrasekharMag2008}
V.~Chandrasekhar, J.~G. Andrews, and A.~Gatherer, ``Femtocell networks: a
  survey,'' \emph{IEEE Comm. Magazine}, vol.~46, no.~9, pp. 59--67, Sept. 2008.

\bibitem{ChandrasekharCDMA2009}
V.~Chandrasekhar and J.~G. Andrews, ``Uplink capacity and interference
  avoidance in two-tier femtocell networks,'' \emph{To appear, {IEEE} Trans. on
  Wireless Comm.}, 2009, [Online] Available at
  \texttt{http://arxiv.org/abs/cs.NI/0702132}.

\bibitem{ZemlianovInfocomm2005}
A.~Zemlianov and G.~De~Veciana, ``Cooperation and decision-making in a wireless
  multi-provider setting,'' in \emph{Proc., IEEE INFOCOM}, vol.~1, Mar. 2005,
  pp. 386--397.

\bibitem{Claussen2007}
H.~Claussen, ``Performance of macro- and co-channel femtocells in a
  hierarchical cell structure,'' in \emph{Proc., IEEE International Symp. on
  Personal, Indoor and Mobile Radio Comm.}, Sept. 2007, pp. 1--5.

\bibitem{Niyato2005}
D.~Niyato and E.~Hossain, ``Call admission control for {Q}o{S} provisioning in
  4{G} wireless networks: issues and approaches,'' \emph{{IEEE} Network},
  vol.~19, no.~5, pp. 5--11, Sept./Oct. 2005.

\bibitem{Ganz1997}
A.~Ganz, C.~M. Krishna, D.~Tang, and Z.~J. Haas, ``On optimal design of
  multitier wireless cellular systems,'' \emph{{IEEE} Communications Magazine},
  vol.~35, no.~2, pp. 88--93, Feb. 1997.

\bibitem{Kishore2005}
S.~Kishore, L.~J. Greenstein, H.~V. Poor, and S.~C. Schwartz, ``Soft handoff
  and uplink capacity in a two-tier {CDMA} system,'' \emph{IEEE Trans. on
  Wireless Comm.}, vol.~4, no.~4, pp. 1297--1301, July 2005.

\bibitem{Klein2004}
T.~E. Klein and S.-J. Han, ``Assignment strategies for mobile data users in
  hierarchical overlay networks: performance of optimal and adaptive
  strategies,'' \emph{IEEE Journal on Sel. Areas in Comm.}, vol.~22, no.~5, pp.
  849--861, June 2004.

\bibitem{Shen2004}
Z.~Shen and S.~Kishore, ``Optimal multiple access to data access points in
  tiered {CDMA} systems,'' in \emph{Proc., IEEE Veh. Tech. Conf.}, vol.~1,
  Sept. 2004, pp. 719--723.

\bibitem{Foschini1993}
G.~J. Foschini and Z.~Miljanic, ``A simple distributed autonomous power control
  algorithm and its convergence,'' \emph{IEEE Trans. on Veh. Tech.}, vol.~42,
  no.~4, pp. 641--646, Nov. 1993.

\bibitem{Zander1992a}
J.~Zander, ``Performance of optimum transmitter power control in cellular radio
  systems,'' \emph{IEEE Trans. on Veh. Tech.}, vol.~41, no.~1, pp. 57--62, Feb.
  1992.

\bibitem{Grandhi1994}
S.~A. Grandhi and J.~Zander, ``Constrained power control in cellular radio
  systems,'' in \emph{Proc., IEEE Veh. Tech. Conf.}, June 1994.

\bibitem{Bambos2000}
N.~Bambos, S.~C. Chen, and G.~J. Pottie, ``Channel access algorithms with
  active link protection for wireless communication networks with power
  control,'' \emph{{IEEE}/{ACM} Transactions on Networking}, vol.~8, no.~5, pp.
  583--597, Oct. 2000.

\bibitem{Zander1992b}
J.~Zander, ``Distributed cochannel interference control in cellular radio
  systems,'' \emph{IEEE Trans. on Veh. Tech.}, vol.~41, no.~3, 1992.

\bibitem{Yates1995a}
R.~D. Yates, ``A framework for uplink power control in cellular radio
  systems,'' \emph{IEEE Journal on Sel. Areas in Comm.}, vol.~13, no.~7, pp.
  1341--1347, Sept. 1995.

\bibitem{Yates1995b}
R.~D. Yates and C.-Y. Huang, ``Integrated power control and base station
  assignment,'' \emph{IEEE Trans. on Veh. Tech.}, vol.~44, no.~3, pp. 638--644,
  Aug. 1995.

\bibitem{Hanly1995}
S.~V. Hanly, ``An algorithm for combined cell-site selection and power control
  to maximize cellular spread spectrum capacity,'' \emph{IEEE Journal on Sel.
  Areas in Comm.}, vol.~13, no.~7, pp. 1332--1340, Sept. 1995.

\bibitem{Ulukus1998}
S.~Ulukus and R.~D. Yates, ``Stochastic power control for cellular radio
  systems,'' \emph{IEEE Trans. on Comm.}, vol.~46, no.~6, June 1998.

\bibitem{Sung2005}
C.~W. Sung and K.~K. Leung, ``A generalized framework for distributed power
  control in wireless networks,'' \emph{IEEE Trans. on Info. Theory}, vol.~51,
  no.~7, pp. 2625--2635, July 2005.

\bibitem{JiHuang1998}
H.~Ji and C.-Y. Huang, ``Non-cooperative uplink power control in cellular radio
  systems,'' \emph{Wire. Net.}, vol.~4, no.~3, 1998.

\bibitem{Goodman2000}
D.~Goodman and N.~Mandayam, ``Power control for wireless data,'' \emph{IEEE
  Personal Comm. Magazine}, vol.~7, no.~2, Apr. 2000.

\bibitem{Saraydar2002}
C.~U. Saraydar, N.~B. Mandayam, and D.~J. Goodman, ``Efficient power control
  via pricing in wireless data networks,'' \emph{IEEE Trans. on Comm.},
  vol.~50, no.~2, Feb. 2002.

\bibitem{Koskie2005}
S.~Koskie and Z.~Gajic, ``A {N}ash game algorithm for {SIR}-based power control
  in 3{G} wireless {CDMA} networks,'' \emph{IEEE Trans. on Networking},
  vol.~13, no.~5, pp. 1017--1026, Oct. 2005.

\bibitem{Xiao2001}
M.~Xiao, N.~B. Shroff, and E.~K.~P. Chong, ``Utility-based power control in
  cellular wireless systems,'' in \emph{Proc., IEEE INFOCOM}, vol.~1,
  Anchorage, AK, USA, 2001, pp. 412--421.

\bibitem{Alpcan2001}
T.~Alpcan, T.~Basar, R.~Srikant, and E.~Altman, ``{CDMA} uplink power control
  as a noncooperative game,'' in \emph{Proceedings of the IEEE Conference on
  Decision and Control}, vol.~1, Orlando, FL, USA, 2001, pp. 197--202.

\bibitem{Altman2006}
E.~Altman, T.~Boulogne, R.~El-Azouzi, T.~Jiminez, and L.~Wynter, ``A survey of
  network games in telecommunications,'' \emph{Computers and Operations
  Research}, pp. 286--311, Feb. 2006.

\bibitem{Ho2007}
L.~T.~W. Ho and H.~Claussen, ``Effects of user-deployed, co-channel femtocells
  on the call drop probability in a residential scenario,'' in \emph{Proc.,
  IEEE International Symp. on Personal, Indoor and Mobile Radio Comm.}, Sept.
  2007, pp. 1--5.

\bibitem{Guvenc2008}
I.~Guvenc, M.-R. Jeong, F.~Watanabe, and H.~Inamura, ``A hybrid frequency
  assignment for femtocells and coverage area analysis for co-channel
  operation,'' \emph{{IEEE} Comm. Letters}, vol.~12, no.~12, Dec. 2008.

\bibitem{Jo2008}
H.-S. Jo, J.-G. Yook, C.~Mun, and J.~Moon, ``A self-organized uplink power
  control for cross-tier interference management in femtocell networks,'' in
  \emph{Proc., Military Comm. Conf.}, Nov. 2008, pp. 1--6.

\bibitem{Choi2008}
D.~Choi, P.~Monajemi, S.~Kang, and J.~Villasenor, ``Dealing with loud
  neighbors: The benefits and tradeoffs of adaptive femtocell access,'' in
  \emph{Proc., IEEE Global Telecomm. Conference}, Nov./Dec. 2008, pp. 1--5.

\bibitem{Qian2007}
L.~Qian, X.~Li, J.~Attia, and Z.~Gajic, ``Power control for cognitive radio ad
  hoc networks,'' in \emph{{IEEE} Workshop on Local \& Metro. Area Netwks.},
  June 2007, pp. 7--12.

\bibitem{Hoven2005}
N.~Hoven and A.~Sahai, ``Power scaling for cognitive radio,'' in
  \emph{International Conf. on Wireless Networks, Communications and Mobile
  Computing}, vol.~1, June 2005, pp. 250--255.

\bibitem{Horn1985}
R.~Horn and C.~Johnson, \emph{Matrix Analysis}.\hskip 1em plus 0.5em minus
  0.4em\relax Cambridge University Press, 1985.

\bibitem{Meyer1989}
C.~Meyer, ``Uncoupling the {P}erron eigenvector problem,'' \emph{Linear Algebra
  and its {A}ppl.}, vol. 114--115, pp. 69--94, Mar. 1989.

\bibitem{Goldsmith}
A.~Goldsmith, \emph{Wireless Communications}.\hskip 1em plus 0.5em minus
  0.4em\relax Cambridge University Press, 2005.

\bibitem{IMT2000}
``Guidelines for evaluation of radio transmission technologies for
  {IMT}-2000,'' \emph{{ITU} Recommendation M.1225}, 1997.

\bibitem{Nash1951}
J.~F. Nash, ``Non-cooperative games,'' \emph{Ann. Math.}, vol.~54, pp.
  289--295, 1951.

\bibitem{Glicksberg52}
I.~L. Glicksberg, ``A further generalization of the {K}akutani fixed point
  theorem with application to {N}ash equilibrium points,'' \emph{Proceedings of
  the American Mathematical Society}, vol.~3, no.~1, pp. 170--174, 1952.

\bibitem{Rosen65}
J.~B. Rosen, ``Existence and uniqueness of equilibrium points for concave
  n-person games,'' \emph{Econometrica}, vol.~33, no.~3, pp. 520--534, 1965.

\bibitem{Debreu52}
G.~Debreu, ``A social equilibrium existence theorem,'' \emph{Proceedings of the
  National Academy of Sciences}, vol.~38, pp. 886--893, 1952.

\bibitem{Chen2007}
J.~K. Chen, T.~S. Rappaport, and G.~de~Veciana, ``Site specific knowledge for
  improving frequency allocations in wireless {LAN} and cellular networks,'' in
  \emph{Proc., IEEE Veh. Tech. Conf.}, Sept./Oct. 2007, pp. 1431--1435.

\end{thebibliography}

\begin{table}[htp]
\caption{System Parameters}
\label{Tbl:SysPrms}
     \centering
     \begin{tabular}{c|  c | c}
     \hline
     \textbf{Variable} & \textbf{Parameter}  & \textbf{Sim. Value} \\ \hline
     $R_c$ & Macrocell Radius  & $1000$ m \\
     $R_f$ & Femtocell Radius    &  $30$ m \\
     $D_{\textrm{grid}}$ &    Grid size   &  $500$ m \\
     $f$ & Carrier Frequency $f_{\textrm{Mhz}}$ & $2000$ MHz \\
     $p_{\textrm{max}}$ & Max. Transmission Power per Mobile & $1$ Watt \\
     $\Gamma_{c,\textrm{min}},\Gamma_{c,\textrm{max}}$ & Max. and Min. Cellular SINR target & $3,10$ dB \\
     $\Gamma_{f,\textrm{min}},\Gamma_{f,\textrm{max}}$ & Max. and Min. Femtocell SINR target & $5,25$ dB \\
     $K_{fi}$ &     Indoor Loss       &  $37$ dB  \\
     $W$ &  Partition Loss  & $5, 10$ dB \\
     $\alpha,\beta$ & Outdoor and Indoor path loss exponents  & $4,3$ \\
     $ t_{\textrm{dB}}$  & Femtocell SINR target reduction & $0.8$ dB \\
     $\delta y$ & Interference threshold reduction & $3$ dB \\
     \hline
\end{tabular}
\end{table}

\begin{algorithm}[htp]
\caption{Maintain cellular link quality at macrocell BS $B_0$}
\label{Al:MaintainLQMacro}
\begin{algorithmic}
    \REPEAT
        \STATE Initialize $k \leftarrow 1, \mathbf{p} \leftarrow \mathbf{p}_{\textrm{max}}$ \COMMENT{Initialize iteration count and TX powers.}
        \WHILE{$k \leq \texttt{MAXITER}$}
            \STATE Cellular user $0$ adapts transmission power according to
           $p_0^{(k+1)} = \min{\left\lbrace\frac{\Gamma_0}{\gamma_0^{(k)}}p_0^{(k)},p_{\textrm{max}}\right \rbrace}$
            \STATE For all $i=1,2,\dots, N$, femtocell user $i$ adapts transmit power according to
                $p_i^{(k+1)}=\min\left\lbrace\frac{p_i^{(k)}}{\gamma_i^{(k)}} \gamma_i^{\ast},p_{\textrm{max}} \right \rbrace$ where $\gamma_i^{\ast} \triangleq \left \lbrack \Gamma_i + \frac{1}{a_i}\ln{\left(\frac{a_i g_{i,i}}{b_i g_{0,i}} \right)} \right \rbrack^{+}$
            \STATE $k \Leftarrow k+1$
        \ENDWHILE
        \STATE Macrocell $B_0$ broadcasts status indicator $\texttt{flag} = \mathbf{1}\lbrack\gamma_0^{\ast} \geq (1-\epsilon)\Gamma_0 \rbrack$ to all femtocells where $\epsilon \in [0,1]$ is a pre-specified tolerance.
        \IF {$\texttt{flag}==0$}
            \STATE \COMMENT{$g_{0,i}$ is channel gain from $B_i$ to $B_0$}
            \STATE Form status indicator at femtocell $B_i$: $\texttt{flag}_i= \mathbf{1} (p_i^{\ast}g_{0,i} > y)$, where $y > 0$
            \IF {$\texttt{flag}_i==1$} \STATE \COMMENT{Reduce reduce $\gamma_i^{\ast}$ since femtocell user $ i$ causes excessive cross-tier Interference.}
                    \STATE SINR Target Update: $\gamma_{i,\textrm{dB}}^{\ast} \Leftarrow \gamma_{i,\textrm{dB}}^{\ast}-t_{\textrm{dB}}$, where $t > 1$
             \ENDIF
             \STATE $y \Leftarrow y/\delta y$
             \COMMENT{Induce more femtocell users to lower SINR Target.}
       \ENDIF
       \COMMENT{Check if cellular user $0$'s SINR is within $(1-\epsilon)\Gamma_0$}
    \UNTIL{$\texttt{flag}==1$}
\end{algorithmic}
\end{algorithm}

\begin{table}[htp]
\caption{Example: Link quality protection for a cellular user (Row 2) with $N=16$ femtocells}
\label{Tbl:Example}
     \centering
     \begin{tabular}{|c | c |  c | c | c | c | c | c|}
     \hline
      \textbf{User $i$}& $\mathbf{D}_{0,i}/\mathbf{R}$ & \textbf{dB Target} $\mathbf{\Gamma}$  & $\mathbf{\Gamma}_{M}^{\ast}$ (dB) & $\mathbf{\Gamma}_{5M}^{\ast}$ (dB) & $\mathbf{\Gamma}_{13M}^{\ast}$ (dB) & $\mathbf{\Gamma}_{19M}^{\ast}$ (dB) & $\mathbf{p}_{19M}^{\ast}$ (dBm)\\ \hline
$0$ & $0.1000$ & $\mathbf{21.0034}$ & $7.8979$ & $9.3358$ & $15.4235$ & $\mathbf{20.1932}$ & $30.0000$ \\
$1$ & $0.2915 $ & $\mathbf{25.3945}$ & $  25.5374 $ & $  25.5374  $ & $ 25.5374 $  & $  \mathbf{23.9538} $ & $   0.4138$ \\
$2$ & $0.1716 $ & $\mathbf{27.8943}$ & $  27.9605 $ & $  27.9605  $ & $ 26.3769 $  & $  \mathbf{21.6260} $ & $   3.1487$ \\
$3$ &$0.1716 $ & $\mathbf{22.6351}$ & $ 22.8535  $ & $ 22.8535  $ & $ 22.8535  $  & $ \mathbf{18.1027}  $ & $ -0.2808$ \\
$4$ &$0.2915 $ & $\mathbf{27.1217}$ & $  27.2182 $ & $  27.2182 $ & $  27.2182 $  & $ \mathbf{ 24.8428} $ & $   1.4084$ \\
$5$ &$0.2506 $ & $  14.0872 $ & $  15.6355 $ & $  15.6355 $ & $  15.6355 $ &  $ 14.8437 $ & $  -3.6491$ \\
$6$ &$0.0850 $ & $\mathbf{14.4560}$ & $  15.3847 $ & $  15.3847 $ & $  10.6339 $  & $  \mathbf{5.8830} $ & $   1.3216 $ \\
$7$ &$0.0850 $ & $\mathbf{ 28.3470} $ & $  28.3891 $ & $  26.8054 $ & $  20.4709 $ & $ \mathbf{15.7201} $ & $  11.1628$ \\
$8$ &$0.2506 $ & $\mathbf{ 25.7148} $ & $  25.8408 $ & $  25.8408 $ & $  25.8408 $ & $ \mathbf{21.8818} $ & $   3.5317$ \\
$9$ &$0.3100 $ & $\mathbf{ 17.9488} $ & $  18.7032 $ & $  18.7032 $ & $  18.7032  $ & $  \mathbf{17.9114}$ & $  -0.5868$ \\
$10$ &$0.2014 $ & $\mathbf{8.4026}$ & $  12.3111 $ & $  12.3111 $ & $  12.3111 $ &  $  \mathbf{7.5602} $ & $   3.0034$ \\
${11}$ &$0.2014 $ & $\mathbf{28.3375}$ & $  28.4014 $ & $  28.4014 $ & $  24.4423 $ &$ \mathbf{19.6914} $ & $  15.1274$\\
${12}$ &$0.3100 $ & $  12.3944 $ & $  14.6515 $ & $  14.6515 $ & $  14.6515 $ &  $  14.6515 $ & $  -3.5588$\\
${13}$ &$0.4301 $ & $   8.6965 $ & $  13.1272 $ & $  13.1272 $ & $  13.1272 $ &  $  13.1272 $ & $ -10.4070$\\
${14}$ &$0.3598 $ & $\mathbf{19.4412} $ & $  20.0152 $ & $  20.0152 $ & $  20.0152 $ & $\mathbf{19.2234}$ & $   0.7828$\\
${15}$ &$0.3598 $ & $\mathbf{20.3513} $ & $  20.8225  $ & $ 20.8225 $ & $  20.8225 $ &  $\mathbf{20.0306}$ & $   1.7930$\\
${16}$ &$0.4301 $ & $   26.7008 $ & $  26.8211 $ & $  26.8211 $ & $  26.8211$ & $  26.8211 $ & $   3.4629$\\
 \hline
\multicolumn{8}{|l|} {User $0$ designates cellular user, while Users $1$ through $16$ represent femtocell users.} \\
\multicolumn{8}{|l|} {Bold faced entries represent either user $0$ or femtocell users unable to meet their SINR target.} \\
\multicolumn{8}{|l|}{The spectral radius $\rho(\mathbf{\Gamma G})=4.4391$, implying that initial SINR targets are infeasible.} \\
\multicolumn{8}{|l|}{Following update $19$ (M=$1000$ iterations/update), the spectral radius $\rho(\mathbf{\Gamma}^{\ast}_{19M}\mathbf{G})=0.9999 < 1$} \\
\hline
\end{tabular}
\end{table}

\end{document}